\documentclass[superscriptaddress,floatfix,amssymb]{
revtex4-2}

\usepackage{graphicx}
\usepackage{bm}
\usepackage{mathtools}
\usepackage{subcaption}
\usepackage{ragged2e}

\usepackage{siunitx}
\usepackage[ruled]{algorithm2e}
\usepackage{multirow}
\usepackage{tikz}
\usetikzlibrary{arrows.meta, positioning}
\usepackage{adjustbox}
\usepackage{colortbl}
\usepackage{amsmath} 

\usepackage[colorlinks]{hyperref}

\usepackage{placeins}
\usepackage{float} 

\begin{document}

\title{
Non-Iterative Disentangled Unitary Coupled-Cluster based on Lie-algebraic structure}

\author{Mohammad Haidar*}
\affiliation{Qubit Pharmaceuticals, Advanced Research Department, 75014, Paris}
\email{\textcolor{blue}{Corresponding authors: Mohammadhaidar2016@outlook.com}}
\email{\newline \textcolor{blue}{alberto.peruzzo@qubit-pharmaceuticals.com}}
\email{\newline \textcolor{blue}{jean-philip.piquemal@sorbonne-universite.fr}}

\author{Olivier Adjoua}
\affiliation{Sorbonne Université, Laboratoire de Chimie Théorique, UMR 7616 CNRS, 75005, Paris, France}
\author{Siwar Badreddine}
\affiliation{Sorbonne Université, Laboratoire de Chimie Théorique, UMR 7616 CNRS, 75005, Paris, France}
\author{Alberto Peruzzo*}
\affiliation{Qubit Pharmaceuticals, Advanced Research Department, 75014, Paris}
\author{Jean-Philip Piquemal*}
\affiliation{Qubit Pharmaceuticals, Advanced Research Department, 75014, Paris}
\affiliation{Sorbonne Université, Laboratoire de Chimie Théorique, UMR 7616 CNRS, 75005, Paris, France}

\maketitle

\section*{Abstract}

Due to their non-iterative nature, fixed unitary coupled cluster (UCC) ansätze are attractive for performing quantum chemistry variational quantum eigensolver (VQE) computations as they avoid pre-circuit measurements on a quantum computer. However, achieving chemical accuracy for strongly correlated systems with UCC requires further inclusion of higher-order fermionic excitations beyond triples increasing circuit depth. We introduce $k$-NI-DUCC, a fixed and non-iterative disentangled unitary coupled cluster compact ansatz, based on specific ``$k$'' sets of ``qubit'' excitations, eliminating the needs for fermionic-type excitations. These elements scale linearly ($\mathcal{O}(n)$) by leveraging Lie algebraic structures, with $n$ being the number of qubits. The key excitations are screened through specific selection criteria, including the enforcement of all symmetries, to ensure the construction of a robust set of generators. NI-DUCC employs ``$k$'' products of the exponential of $\mathcal{O}(n)$- anti-Hermitian Pauli operators, where each single Pauli string has a length $p$. This results in a fewer two-qubit CNOT gates circuit scaling, $\mathcal{O}(knp)$,  suitable for hardware implementations. Tested on LiH, H$_6$ and BeH$_2$, NI-DUCC-VQE achieves both chemical accuracy and rapid convergence even for molecules deviating significantly from equilibrium. It is hardware-efficient, reaching the exact full configuration interaction energy solution at specific layers, while reducing significantly the VQE optimization steps. While NI-DUCC-VQE effectively addresses the gradient measurement bottleneck of ADAPT-VQE-like iterative algorithms, the classical computational cost of constructing the $\mathcal{O}(n)$ set of excitations increases exponentially with the number of qubits. We provide a first implementation for constructing the generators' set, able to handle up to 20 qubits and discuss the efficiency perspectives.

\section*{\textbf{Introduction}}

Quantum chemistry is one of the most promising fields for the application of early quantum computing   \cite{aspuru2005simulated,reiher2017elucidating,bauer2020quantum,bassman2021simulating}.
Quantum algorithms specifically designed for quantum chemistry include, for example, quantum phase estimation (QPE) \cite{kitaev1995quantum}, variational quantum eigensolvers (VQE)  \cite{peruzzo2014variational}, and quantum imaginary time evolution (QITE) among others \cite{mcardle2020quantum, cao2019quantum, bauer2020quantum, yeter2021benchmarking}. The current landscape of quantum computers resides within the noisy intermediate-scale quantum (NISQ) era, which is constrained primarily by the number of qubits and by the circuit depth \cite{whitfield2011simulation, preskill2018quantum, elfving2020will, bharti2022noisy}. Within this context, the VQE family of techniques emerges as a particularly promising avenue applicable to NISQ hardware \cite{mcardle2020quantum, peruzzo2014variational, mcclean2016theory, mcardle2020quantum, cerezo2021variational}. It has demonstrated successful execution on real NISQ quantum computers, employing various technologies such as superconducting qubits \cite{google2020hartree, kandala2017hardware, robledo2024chemistry}, photons \cite{peruzzo2014variational,maring2024versatile}, and trapped ions \cite{hempel2018quantum,ollitrault2024estimation}. 
Indeed, VQE is a hybrid quantum-classical algorithm aimed at solving the eigenvalue problem, a problem found in many fields.\cite{peruzzo2014variational} Chemistry was its first application and VQE has been used at determining the ground-state energy of a given Hamiltonian.

Therefore, a crucial element of VQE is the design of the parameterized ansatz. One initial approach in VQE relies on the use of the unitary coupled cluster singles and doubles (UCCSD) approach \cite{romero2018strategies,sokolov2020quantum,xia2020qubit,anand2021quantum}. It provides a well-defined fermionic excitation hierarchy, while maintaining size consistency and exhibiting variational properties. Furthermore, it achieves high accuracy and rapid convergence for near-equilibrium molecules, as demonstrated by successful experimental tests on small molecules \cite{peruzzo2014variational,shen2017quantum,o2022purification,Benzenewassil}. However, the non-commutativity among operators makes it challenging to directly translate the UCCSD ansatz into quantum gates, requiring the use of an additional Trotter approximation. 
In practice, this non-commutativity introduces complexities in operator ordering, which can significantly impact the resulting energy variance and potentially lead to non-uniqueness in the resultant wavefunction \cite{grimsley2019trotterized}. Moreover, as the molecular size increases, Trotterized UCCSD excitations  may fail  to achieve  chemical accuracy while incorporating numerous unnecessary excitations. These limitations affect practical implementations on real hardware, as the increased circuit depth introduces additional noise \cite{Benzenewassil}. 
Incorporating higher-order excitations remains possible to effectively capture the missing correlation effects of the Trotterized UCCSD ansatz. To address this, methods such as Trotterized Symmetry unitary coupled cluster singles, doubles, triples (Sym-UCCSDT) and unitary selective coupled cluster singles, doubles, triples, quadruples (UsCCSDTQ) have been proposed.  Sym-UCCSDT incorporates both spin and orbital symmetries \cite{haidar2023extension}, while UsCCSDTQ uses selected CI determinants \cite{fedorov2022unitary}.  Both methods aim to select only the necessary excitations, improving chemical accuracy and accelerating the optimization process in VQE for larger molecules.
However, while they provide improvements upon the native Trotterized UCCSD ansatz, they also mechanically increase the circuit depth through the inclusion of higher-order excitations.

To address these issues, constructing an ansatz tailored to system-specific criteria offers a promising solution for achieving shallower circuits.
The popular ADAPT-VQE iterative algorithm is a notable example of this strategy \cite{grimsley2019adaptive}. It effectively extracts electronic correlation energy using a small set of operators. However, its operator selection results in high computational overhead due to gradient measurements. These measurements are essential for evaluating the impact of adding candidate operators to the ansatz. At each iteration, the algorithm calculates the energy gradient, indicating how much the energy changes with each addition. This process is expensive and leads to increasing overhead as the ansatz grows. Despite subsequent efforts to minimize these costs, ADAPT-VQE variants \cite{tang2021qubit,liu2021efficient,yordanov2021qubit,anastasiou2022tetris,shkolnikov2023avoiding,feniou2023adaptive,feniou2023overlap,traore2024shortcut} continue to face gradient measurement challenges, especially as molecular size increases, while maintaining the necessary chemical accuracy. Similar to ADAPT-VQE methods, the Qubit coupled cluster (QCC) approach \cite{ryabinkin2018qubit,ryabinkin2018constrained} and its variants, including iterative Qubit coupled cluster (iQCC) \cite{ryabinkin2020iterative}, offer promising strategies for designing a compact wavefunction.
In iQCC, generators are chosen from a specific set of operators, including all possible Pauli strings, to ensure energy reduction. These generators are sampled using first-order derivative calculations, resulting in a shallow circuit depth. Unlike ADAPT-VQE, the iQCC method avoids the need for full re-optimization of all parameters at each iteration, enhancing its practicality for NISQ devices.
However, this advantage comes with costs: each iteration requires a canonical transformation of the Hamiltonian, which leads to an expanded Hamiltonian size. In practice, this requires high computational resources on classical computers, coupled to additional measurements on quantum devices, and gradient calculations at each step. Furthermore, the iQCC method does not fully guarantee the ordering of operators\cite{ryabinkin2020iterative}, which can increase the risk of encountering local minima and contribute to the ‘barren plateaus’ problem \cite{cerezo2021cost}. 
Nevertheless, recent advances have been made in exploring efficient techniques for QCC methods, such as in \cite{lang2020unitary,lang2023growth}. In addition to the qubit coupled cluster (QCC) and ADAPT-VQE approaches, recent advancements such as the tiled unitary product state (tUPS) \cite{burton2024accurate},  the COMmutativity Pre-screened Automated Selection of Scatterers (COMPASS) \cite{mondal2023development}, the Cognitive Optimization via Many-body Perturbation theory towards Ansatz Construction and Tailoring (COMPACT) \cite{halder2024noise}, and subspace projective quantum eigensolver (SPQE) \cite{patra2024towards} methods have significantly improved computational accuracy and the efficiency of quantum computing resources.

In this paper,  we introduce the non-iterative disentangled unitary coupled cluster (NI-DUCC) method, a novel and efficient approach designed for NISQ devices. NI-DUCC simplifies the unitary coupled cluster framework by operating  directly in qubit space excitations,  that are optimally generated through Lie algebraic structures. These operators (each of length $p$) are anti-Hermitian and their set scales linearly with the number of qubits, $n$,  which significantly reduces the reliance on error-prone two-qubit CNOT gates (see ``Quantum computer datasheet'', Google, 2021 \cite{google_quantum_ai_datasheet}). 
The NI-DUCC-VQE algorithm avoids gradient measurements and integrates symmetries to accelerate convergence and optimization.  Additionally, it  leverages Lie algebraic relations to overcome ordering issues typically seen in Trotterized UCC methods. This may explain why our simulations across various molecules show that NI-DUCC effectively reduces VQE energy without barren plateau issues \cite{Ragone1}. While also ADAPT-VQE is considered relatively immune to barren plateaus and local minima as shown in some simulations \cite{grimsley2023adaptive}, it still faces challenges with gradient troughs that can hinder smooth convergence. In contrast, NI-DUCC achieves consistent VQE energy reduction, offering a smoother and more reliable convergence path. Furthermore,  it achieves Full Configuration Interaction (FCI) solutions as the number of layers ($k$) increases, requiring significantly fewer optimization steps, and it scales CNOT gate counts as $\mathcal{O}(knp)$
 per circuit, which makes it a promising approach for current quantum hardware.

The rest of the paper is organized as follows. The first section provides an introduction to the NI-DUCC ansatz, followed by a detailed explanation of the NI-DUCC-VQE algorithm. In Section 2, we describe the computational methods and present various numerical tests on several molecules. To showcase the NI-DUCC-VQE capabilities, especially in achieving chemical accuracy, we compare its results to those obtained from UCCSD, UCCSDT, sym-UCCSDT, UsCCSDTQ, COMPASS, COMPACT, ADAPT-VQE, and to the reference FCI classical approach. In the final section, we examine the NI-DUCC resource requirements and compare its performance to that of current state-of-the-art methods.

\section{\textbf{Methodology}}
 
\subsection{Disentangled unitary coupled cluster (DUCC): Lie Algebraic structure and symmetric excitations}

Instead of adhering to the fermionic picture as in the UCCSD \cite{grimsley2019trotterized}, UCCGSD \cite{lee2018generalized}, $k-$UpCCGSD \cite{lim2022quantum} techniques, we propose to develop an approach based on the ``Disentangled'' unitary coupled cluster (DUCC) method introduced by Evangelista et al.  \cite{evangelista2019exact}. This involves using operators that act directly within the multi-qubit space, derived from a pre-selected set of operators of size $M \in \mathbb{N}^*$:
\begin{equation}
\displaystyle
\label{disentangled}
\hat{U}(\vec{\theta)} = \prod^M_{l=1} e^{i \vec{\theta}_l \Hat{P_l}},
\end{equation}
where $\Hat{P_l}$ is the so-called Pauli string operator, formed by the tensor product of Pauli matrices (with $l \in  \left\{1,\ldots,M \right\}$),  and $\vec{\theta}=\{\vec{\theta_l}\}_{l=1}^{M}$ are the variational parameters to be optimized. These operators do not necessarily commute, which can lead to a disordered wavefunction and significantly impact the results\cite{Grimsleytrotter}. To address this ordering issue, we review the Lie algebraic properties of a specific set of operators below, ensuring they satisfy the closure relations. Moreover, since the main goal of the present work is to develop a compact ansatz suitable for NISQ devices, it is ideal to keep the set of operators as small as possible. Ideally, it should grow linearly  with the number of qubits, $n$. In practice, this can be achieved by applying symmetries to the multi-qubit space operators, which facilitates the optimization of  the generators set. These symmetries include the particle number $\hat{N_e}$, spin symmetries ($\hat{S}^2 = 0$, $\hat{S_z}$ = 0), and symmetry point group (or spatial symmetry $ \hat{R}$). We refer to these operators as ``starters'' since they act directly on top of the Hartree-Fock state, guaranteeing a symmetric initial wavefunction.  This is, in turn, expected to accelerate convergence, as demonstrated in previous studies  \cite{shkolnikov2023avoiding,gard2020efficient}. 
Assume now we have a set of starters $\mathcal{S}_j = \{\hat{P}_l\}_{l=1}^{M_s}$ of length $M_s$, with $\hat{P}_l$ being symmetry-preserving operators. Then, it would be possible to build a sub-algebra from this set, denoted as $\mathcal{S}$, involving $M \le 4^{n-1}$, satisfied by closure relations such that for any pair 

\begin{equation}
\displaystyle
    [\hat{P}_i, \hat{P}_j] = \sum_{k} c(k)_{ij} \hat{P}_r ,\,\,\, 
\end{equation}
with $\hat{P}_i, \hat{P}_j, \hat{P}_r \in \mathcal{S}$.  
In practice, utilizing a set $\mathcal{S}$ that grows exponentially with 
$n$ is impractical, as it would compromise the efficiency  of using quantum computers to represent $\hat{U}(\vec{\theta})$. Instead, we aim to reduce the set of excitations into a size that corresponds to a linear function of 
$n$. To do so, we leverage the advantage of retaining the set of starters $\mathcal{S}_j$, as they effectively  help in lowering the energy and enable a faster optimization. Then, it requires to find a set $\mathcal{S}_{j+1}$ with an optimal number of $\hat{P}_l$  such that the union $\mathcal{S}^{(c)} = \big\{\mathcal{S}_j \cup \mathcal{S}_{j+1}\big\}$ generates a Lie sub-algebra of length $\ll  4^{n-1}$. To determine the optimal size of
 $\mathcal{S}_{j+1}$, we reviewed the findings from reference \cite{shkolnikov2023avoiding}. It was demonstrated that a minimum complete pool (MCP) of $2n - 2$ Pauli operators strings is sufficient for achieving chemical accuracy with qubit ADAPT-VQE for molecules such as H$_4$, LiH, and BeH$_2$.  Following the guidelines for constructing an MCP proposed in reference \cite{shkolnikov2023avoiding}, we generate excitations in the NI-DUCC approach with the set  $\mathcal{S}^{(c)} = \big\{\mathcal{S}_j \cup \mathcal{S}_{j+1}\big\}$, ensuring that its size is $2n - 2$: 
\begin{equation}
\displaystyle
    \label{disentangled1}
\hat{U}(\vec{\theta}) = \prod^{2n-2}_{l=1} e^{i \vec{\theta}_l \hat{P_l}}.
\end{equation} 

As a result, the above equation represents a ``compact'' and ``ordered'' DUCC wavefunction, relying heavily on a sufficient number of starters, $\mathcal{S}_{j}$, that adhere to all symmetries. However, the ``strength'' of these symmetric starters significantly influences convergence, as demonstrated in our simulations (see Section (\ref{numerical})). This strength is measured using fermionic pre-screening excitations, as detailed in the algorithm below. 
Given that the NI-DUCC wavefunction operates directly within the space of multi-qubit operators, $\hat{P_l}$,  with minimal complexity of two-qubit CNOT gates, it can be tailored to fit specific hardware constraints. In this line, the NI-DUCC approach can be connected to the hardware efficient ansatz \cite{gard2020efficient}, which employs multiple layers of symmetry-preserving gates with various variational parameters. Thus we can similarly implement the NI-DUCC wavefunction by incorporating $k$ layers, each consisting 
of $2n-2$ operators (see step 2 in Figure 1).

The resulting $k$-NI-DUCC wavefunction can be expressed as:
\begin{equation}
\displaystyle
    \label{disentangled_comp}
|\Psi(\vec{\theta})\rangle  = \biggl[\,\,\prod^{k}_{m=1}\bigg(\prod^{2n-2}_{l=1} e^{i \vec{\theta}_l \hat{P}_l}\bigg)\,\,\biggr]|\psi_{HF}\rangle.
\end{equation} 

This wavefunction offers flexibility as it allows users to adjust the depth of layers $k$ in the circuit. As shown in the results section, increasing the number of layers improves accuracy. If a certain level of layers is added, it even allows to reach the FCI energy solution, with fewer optimization steps.
For the systems studied  (detailed in Section \ref{costs}), k-NI-DUCC consistently reached FCI solution with $k=8$ layers, demonstrating efficient scaling in capturing correlations. We found that the NI-DUCC ansatz outperformed existing methods in CNOT gate efficiency but it requires 2-3 times more parameters than ADAPT-VQE, depending on the system.
In terms of general applicability, while $k$ can sometimes be roughly estimated based on system size and bond length deviations, it is typically more reliable to iteratively test performance across values of $k$ to identify the optimal layer count for a given accuracy. This iterative approach helps ensure efficient resource use in terms of parameters and CNOT gates, particularly for larger  molecular systems where expressivity demands can significantly vary.
In addition to its flexible structure, NI-DUCC bypasses the need for iterative energy gradient calculations, a major bottleneck in ADAPT-VQE (we refer the reader to Table S1 in the supplementary materials \cite{thisSuppl} for the scaling of CNOT gates, Parameter counts and residual gradients across different versions of ADAPT-VQE). 

\subsection{The NI-DUCC-VQE protocol}
\label{DUCC-VQE algorithm}
In this section, we  discuss the 4 preparatory components of  NI-DUCC-VQE algorithm.  We refer the reader to section 2 of the Supplementary Materials \cite{thisSuppl}, which systematically describes the basic outline of the algorithm. 
First, we compute the one- and two-electron integrals denoted as {$h_{ij}$ (Eq. 2 of the supplementary material \cite{thisSuppl})} and $v_{ijkl}$ (Eq. 3 of the Supplementary Materials \cite{thisSuppl}), which is a task that can be efficiently performed on a classical computer within polynomial time. Subsequently, we calculate the electronic structure Hamiltonian (Eq. 1 of the Supplementary Materials \cite{thisSuppl}), a standard step in every VQE algorithm. 
Second, we generate all double excitations required to construct the selective UsCCD ansatz, which is a simplified version of UsCCSDTQ \cite{fedorov2022unitary}. We denote these excitations by the $(i,j,k,l)$ indices and add them to a list, denoted $D$. We then initialize an array of zeros for the parameters associated with these double excitation operators, which serve as the initial guess for the UsCCD-VQE process.
For the third component, we select the Hartree-Fock state $|\psi_{HF}\rangle$, as the initial reference state.
Then our objective is to construct the unitary operator described in Eq. (\ref{disentangled1}), thereby completing the NI-DUCC wavefunction (Eq. \ref{disentangled_comp}). To accomplish this, we will outline the following three steps, each visualized in the schematic of the NI-DUCC-VQE algorithm in Figure 1 below:
\begin{enumerate}
\item \textbf{Selection of the double excitations-type fermionic operators}. 
We define  a threshold list $\epsilon \in 10^{-m}$, where $m \in \mathbb{N}^*$, and select all orbitals $(i,j,k,l)$, such that $v_{ijkl} \ge \epsilon$, with $\epsilon = 10^{-2}$.
This selection process helps to construct the fermionic sparse UsCCD operators. We can further identify sparse double excitation operators by using UsCCD-VQE. In each subsequent VQE, we initialize the parameters $t_{ijkl}$ with zero guesses as mentioned above, then optimize them to obtain the optimized parameters, denoted as $t_{ijkl}^*$.  Then we select the excitations whose multiplication values, expressed as $v_{ijkl}.t^*_{ijkl}$, exceed a threshold $\epsilon = 10^{-3}$. We then store the optimized parameters $t_{ijkl}^*$  for subsequent iterations to incorporate more dominant excitations. 
Such a technique corresponds to fermionic pre-screening criterion and is inspired from previous works  \cite{holmes2016heat,li2018fast}: $|v_{ijkl}.t^*_{ijkl}| \ge \epsilon$, which has been performed in the context of UsCCSDTQ-VQE simulations \cite{fedorov2022unitary}.

\item \textbf{Refining the selection of double excitations through symmetries enforcement.} 
To obtain the qubit Pauli string operators needed for constructing the NI-DUCC real wavefunction (see Eq.\ref{disentangled}), we create an odd Pauli string, $\hat{P}$, from each of the pre-selected fermionic operator $(i,j,k,l)$, which involves an odd number of $Y$ Pauli matrices. We then filter out the Pauli strings that violate the symmetry rules when acting on the Hartree-Fock state, such as $\Hat{N_e}$, $\Hat{S}^2$, $\Hat{S}_z$ as well as the Abelian point group operator $\Hat{R}$ (the reader should refer to the detailed examples of such operations presented in section 4 of the Supplementary Materials.\cite{thisSuppl}). The use of this technique ensures that the Pauli operators satisfy the condition $\displaystyle \langle \psi_{HF}|[\Hat{H},\Hat{P}_l]|\psi_{HF}\rangle\ne 0$. 
This means that the operators $\Hat{P}_l$ are selected as initial operators because their commutator with the Hamiltonian is significantly non-zero. These excitations are of double-type in nature. Since the molecular Hamiltonian contains only single and double excitations, any excitations that create more than double excitations on top of the Hartree-Fock state would result in zero commutators. We denote the operators $\Hat{P}_l$ as strongly symmetric qubit excitations.

\item \textbf{Forming the  NI-DUCC wavefunction upon  Lie algebraic closure relations.}
Following the guidelines outlined in reference  \cite{shkolnikov2023avoiding}, we generate a symmetry-preserving MCP of excitations from the strongly symmetric qubit excitations ($\Hat{P}_l$) obtained in the previous step, ensuring a closed Lie algebraic relation (see section 4 of the Supplementary Materials \cite{thisSuppl} for the rules on constructing such symmetry-preserving MCP). These excitations are then used to construct the NI-DUCC wavefunction.
\end{enumerate}

\begin{figure}[h]
  \label{schematics}
\centering
\includegraphics[width=\textwidth]{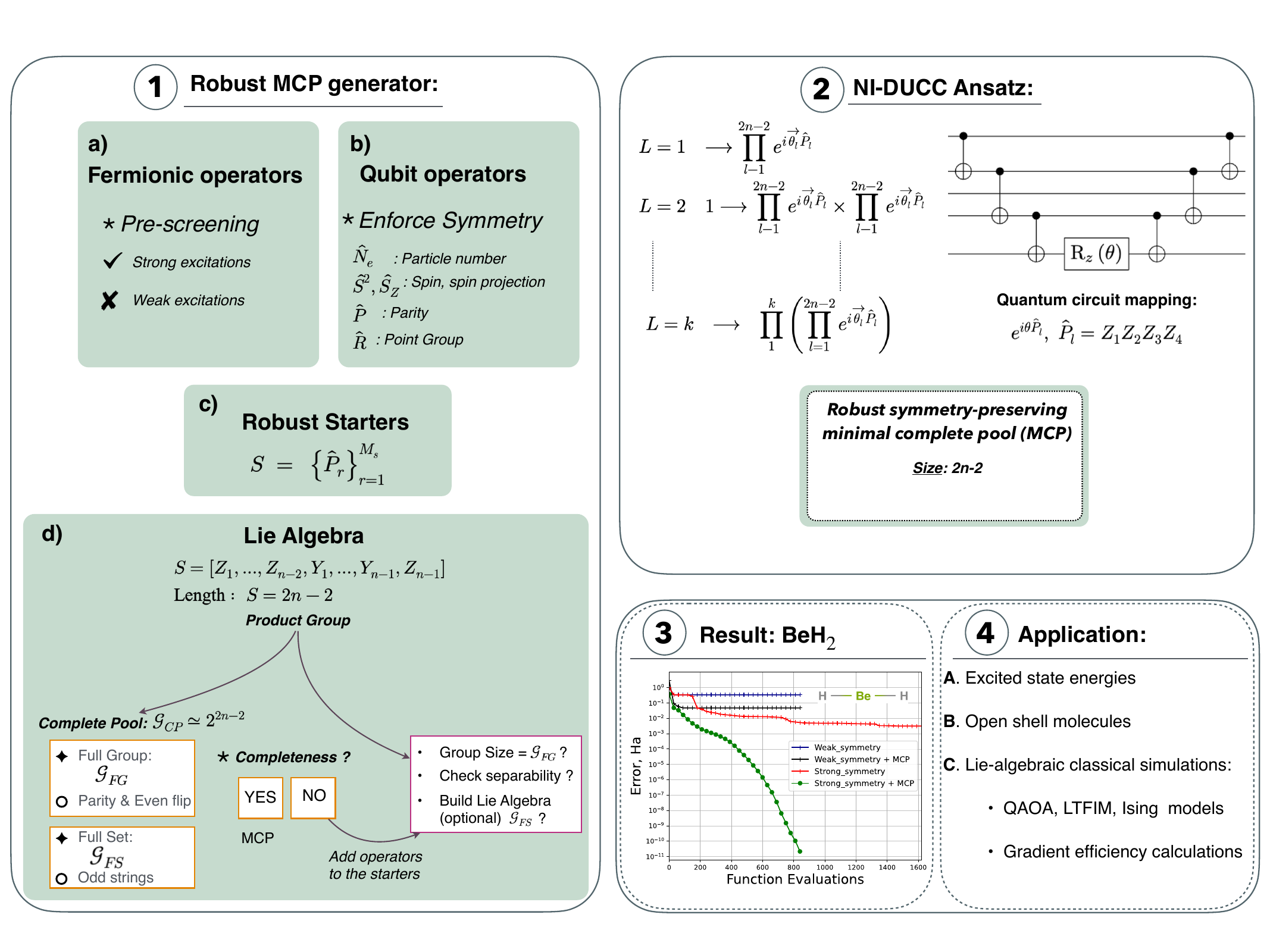}
\caption{\RaggedRight{\textbf{Schematic of NI-DUCC-VQE algorithm}}. \textbf{(1)} The necessary conditions for constructing a robust symmetry-preserving minimal complete pool (MCP)  of size $2n-2$. \textbf{(1.c)}  The initial qubit excitations (starters) are strong, symmetric operators, selected via a fermionic pre-screening criterion based on a threshold $\epsilon$, ensuring only dominant excitations are included while weaker excitations are filtered out. \textbf{(1.d)} Verifying the completeness of the set of the chosen starters (More details can be found in section 4 of  the Supplementary Materials  }
 \cite{thisSuppl}).
\textbf{(2)} Once the MCP is generated, the NI-DUCC ansatz is constructed from  $k$ layers. To the left,  an example of mapping the exponential of a Pauli operator into a circuit is shown, by using staircase method \cite{mcardle2020quantum}. \textbf{(3)} Numerical results tested on BeH$_2$ (14 qubits), using NI-DUCC-VQE, showing the comparison between  robust and non-robust MCP excitations, in terms of chemical accuracy. \textbf{(4)} A list of applications that can benefit from NI-DUCC-VQE and  its ansatz. For further details on extending the NI-DUCC method to excited states or open-shell molecules, including techniques for carefully enforcing symmetries, please refer to Section 4.A in the Supplementary Materials\cite{supplementary}.
\label{figure1}
\end{figure}
\section{\textbf{Computational Details}}
\label{computational details}

Calculations for this work were performed using an initial set of Python development codes. 
The Pyscf package \cite{sun2018pyscf} provided some essential quantities required to perform the NI-DUCC-VQE steps including the one-body $h_{ij}$ and two-body integrals $v_{ijkl}$ (see their explicit equations in section 1 of the Supplementary Materials \cite{thisSuppl}).
UsCCD wavefunction calculations were performed without using Trotterization approximations. We used  the \textit{expm\_multiply} function from
the SciPy library \cite{2020SciPy-NMeth} to compute the action of the matrix exponential of fermionic excitations $T-T^{\dagger}$ on the $|HF\rangle$. The same function was employed to compute the NI-DUCC wavefunctions,  when Qubit-Pauli operators ($\Hat{P}_{k,s}$) are used instead of fermionic $T-T^{\dagger}$  representation (see Eq. \ref{disentangled} and Eq. \ref{disentangled1}). 

To generate the symmetry-preserving MCP excitations, which is the final step of the fourth component needed for the completion of the NI-DUCC algorithm, we used an additional python code, that generates an MCP by ensuring that the pool of excitations meets the required completeness conditions. This involves computing the product group and building their corresponding Lie algebra (see the python code given in \cite{VladShkolnikov}). It is important to note that computing the product group of a given pool is  resource-intensive, especially as the number of qubits increases. Consequently, Section \ref{challenges} provides detailed information about the algorithm computational resource demands while increasing the number of qubits. Finally, to address this issue in order to better estimate the computational requirements, we converted completely the python code into a more efficient C++ capable of a faster MCP generation.

PySCF  \cite{sun2018pyscf} was also used to perform the following steps: (i) initiate molecular geometries and (ii) compute the classical chemistry methods, Hartree-Fock and FCI.  We used the gradient-based optimization method BFGS from the \textit{scipy.optimize} library  \cite{virtanen2020scipy} to optimize the variational parameters (parameters are given in Eq. \ref{disentangled1}). To enhance the optimization speed, we used a BFGS optimizer with an analytically calculated energy gradient vector with a gradient norm of $10^{-10}$ Hartree.  For all simulations, we computed the molecular Hamiltonians using the Slater-type orbital-3 Gaussians (STO-3G) minimal basis set without assuming frozen orbitals.
  
To demonstrate the performance of NI-DUCC-VQE in terms of accuracy and computational resources, we compared it with other adaptive quantum computing methods such as: Fermionic-ADAPT-VQE \cite{grimsley2019adaptive}, Qubit-ADAPT-VQE \cite{tang2021qubit}, QEB-ADAPT-VQE \cite{yordanov2021qubit}, and UsCCDTQ \cite{fedorov2022unitary}. For the ADAPT-VQE methods, we extracted all the numerical data from Yordanov's calculations available at \cite{JordanovSJ}. However, we calculated the UsCCSDTQ-VQE by extending the UsCCD ansatz with single, triple, and quadruple fermionic excitations, implementing them by using the integrated myqlm-fermion/OpenVQE package  \cite{haidar2022open} tools  (such as fermionic second quantization of $\hat{H}$ for a given molecule, etc). 
In the present work, we use up to 24 CPUs cores (2 x Intel(R) Xeon(R) Silver 4116 CPU @ 2.10GHz 12 Cores) at the Open-MP level (see Supplementary Materials \cite{thisSuppl} for further discussion).

\section{NI-DUCC-VQE}
\subsection{Numerical Simulations}
\label{numerical}

Numerical simulations were performed on the H$_6$ and BeH$_2$ molecules, to compare the use of weak and strong symmetric excitations,  both with and without incorporating Lie algebraic closure relations in  the construction of NI-DUCC ansätze. The H$_6$ and BeH$_2$ molecules are frequently simulated in quantum chemistry to benchmark VQE protocols \cite{grimsley2019adaptive,xia2020qubit,yordanov2021qubit,tang2021qubit,shkolnikov2021avoiding,fedorov2022unitary,haidar2023extension,halder2024noise}. These prototypical molecular systems are good model for the evaluation of strongly correlated ground states.
The STO-3G computations for  H$_6$ and BeH$_2$ required the use of 12 and 14 spin-orbitals that were  represented by 12 and 14 qubits respectively.

To study the NI-DUCC-VQE performance on these molecules, we compare various types of excitation evolutions, based on their effectiveness in constructing the NI-DUCC ansatz: (i) \textit{weakly} symmetric excitations; (ii) \textit{weakly} symmetric excitations as starters incorporating \textit{Lie algebraic closure} relations; (iii) \textit{strongly} symmetric excitations; and (iv) \textit{strongly} symmetric excitations as starters incorporating  \textit{Lie algebraic closure} relations.
Furthermore, to ensure a fair comparison among the four protocols, we set the number of layers equal to $k=8$ (see  Eq. \ref{disentangled_comp}).  Then we considered the following cases: 

\begin{enumerate}
    \item the protocols that assign \textit{weak and strong} symmetric excitations use the same number of operators.
    \item the protocols that assign \textit{weak and strong} symmetric excitations as well as  incorporating \textit{Lie closure} relations, use the same number of operators. In that case, the equality in the number of operators is ensured thanks to  the symmetry-preserving MCP generation process, based on Lie algebraic structures, that produces $2n-2$ excitations (see discussion on the previous section).
\end{enumerate}

\begin{figure}
     \centering
    \begin{minipage}[b]{0.65\linewidth}
        \centering
        \includegraphics[width=\linewidth]{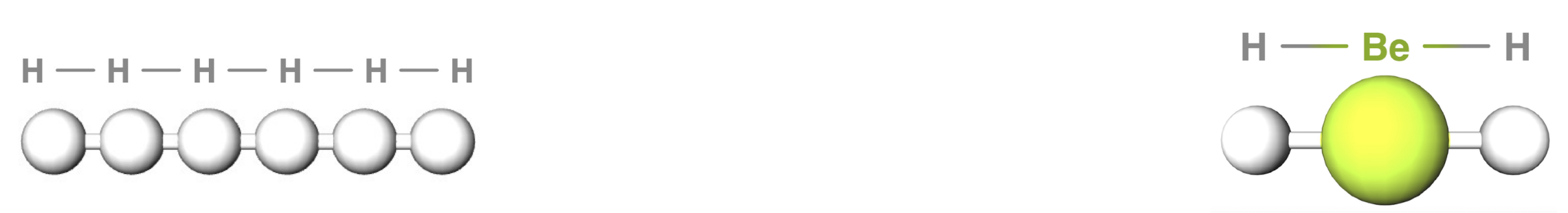}
        \label{fig:mole}
    \end{minipage}
    \begin{subfigure}[b]{0.45\textwidth}
        \centering
        \includegraphics[width=\textwidth]{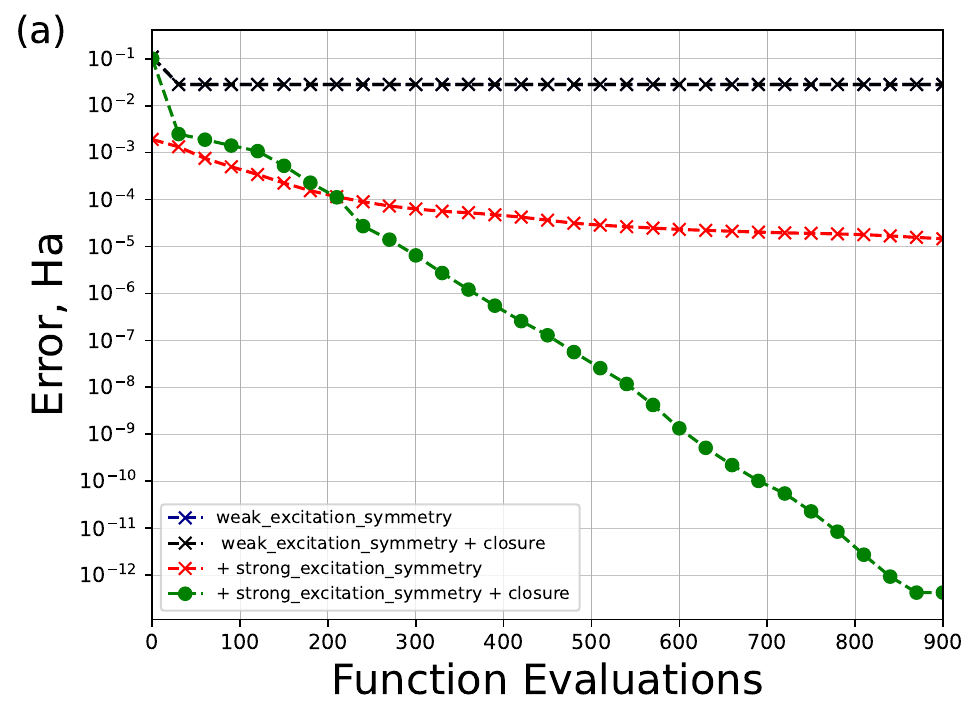}
        \label{fig:sub1}
    \end{subfigure}
     \begin{subfigure}[b]{0.45\textwidth}
        \centering
        \includegraphics[width=\textwidth]{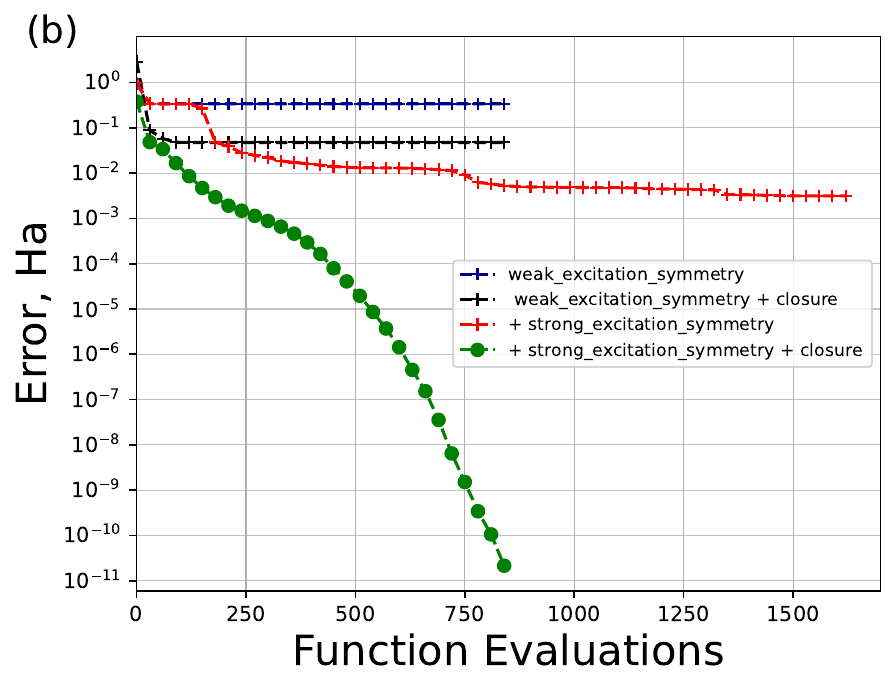}
\captionlistentry{}
        \label{fig:sub3}
    \end{subfigure}
    \vspace{0.5cm}
    \begin{subfigure}[b]{0.45\textwidth}
        \centering
        \includegraphics[width=\textwidth]{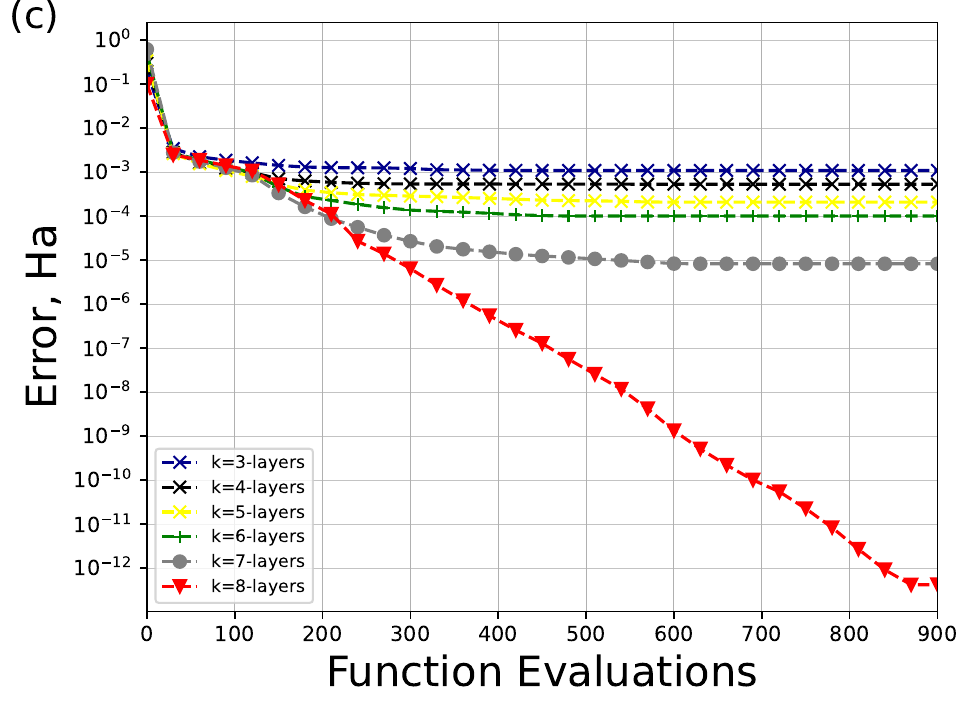}
     \captionlistentry{}
        \label{fig:sub2}
    \end{subfigure}
     \begin{subfigure}[b]{0.45\textwidth}
        \centering
        \includegraphics[width=\textwidth]{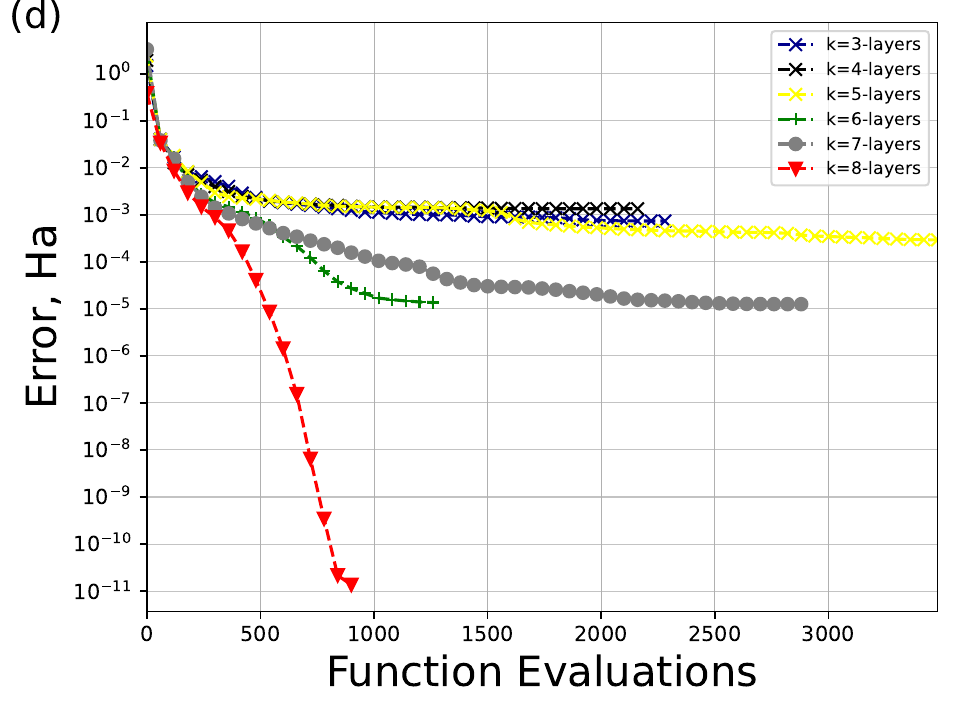}
       \captionlistentry{}
        \label{fig:sub4}
    \end{subfigure}
     \begin{subfigure}[b]{0.45\textwidth}
        \centering
        \includegraphics[width=\textwidth]{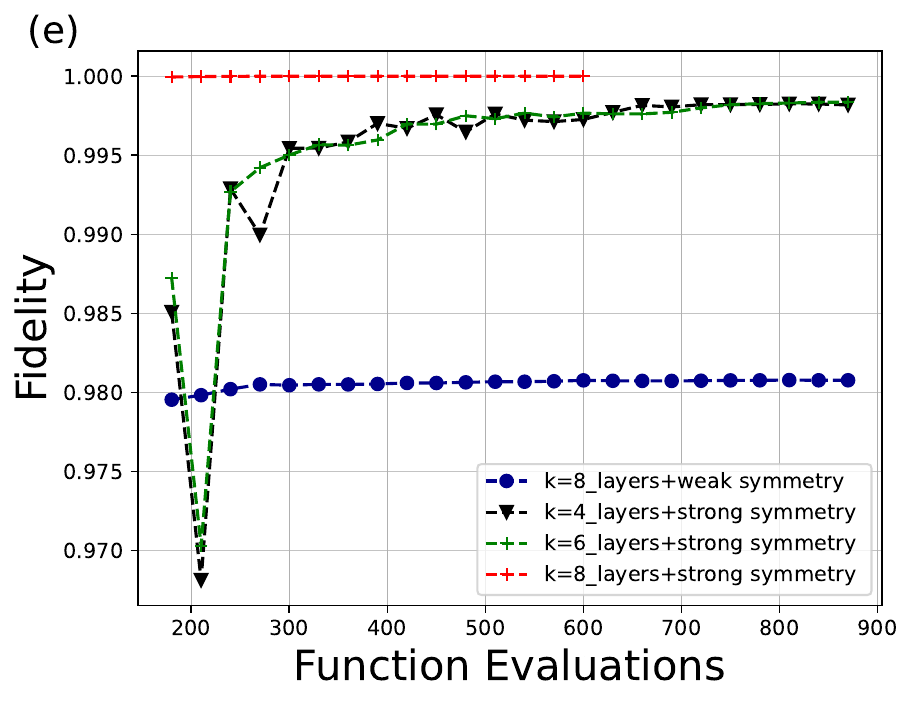}
      \captionlistentry{}
        \label{fig:sub5}
    \end{subfigure}
    \begin{subfigure}[b]{0.44\textwidth}
        \centering
        \includegraphics[width=\textwidth]{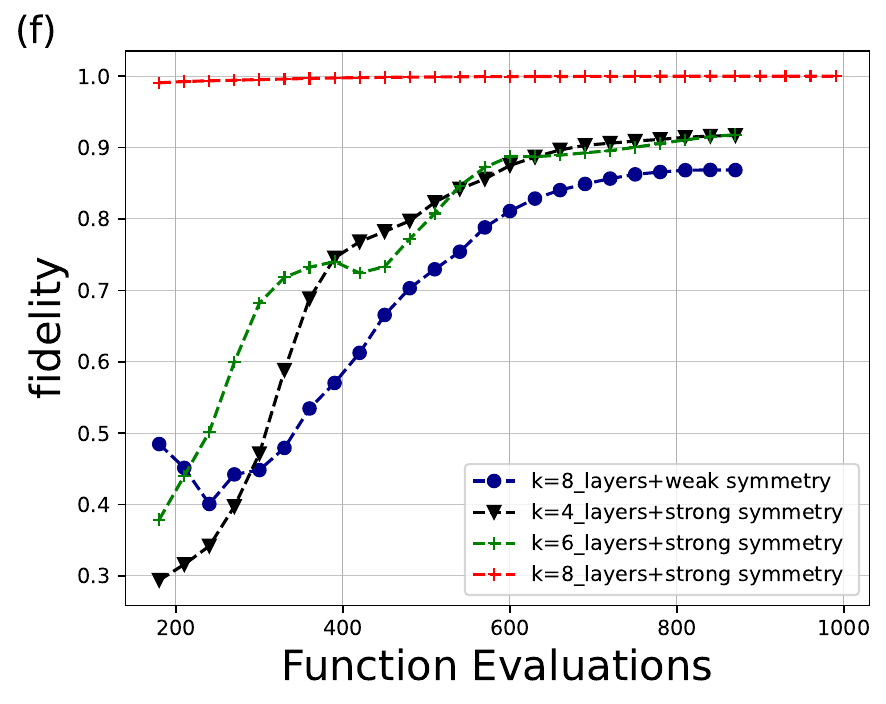}
       \captionlistentry{}
        \label{fig:sub6}
    \end{subfigure}
    \caption{\RaggedRight
    Energy convergence plots for ground states of  H$_6$ (12 qubits)  and BeH$_2$ (14 qubits), using the STO-3G basis  set, at bond distances ( $r_{\text{H-H}}$  = 1.0 \AA,) and ($r_{\text{Be-H}}$ = 3.5 \AA), respectively. The plots are obtained using the NI-DUCC-VQE algorithm: Top panel \textbf{(a, b)}, compares the \textbf{quality of excitation set}, insuring the significance of strong symmetric excitations when they are combined with the closure Lie algebraic properties. In Figure 2(a), the weak symmetric excitations with or without the Lie algebraic closure are overlapping. Panel \textbf{(c, d)}, illustrates the \textbf{rapid convergence} as the number of layers, $k$, increases in the circuit. Bottom Panel \textbf{(e, f)},  shows the  \textbf{fidelity} property, which is calculated by  overlapping $\langle\Psi_j(\vec{\theta^*}) |\Psi_{g}\rangle$ between the  computed NI-DUCC state at each optimization step ``$j$'' and the theoretical  eigenvector ground state $|\Psi_{g}\rangle$ of Hamiltonian ($\hat{H}$).}
    \label{fig:main1}
\end{figure}

The energy convergence plots for the ground states of H$_6$ (Figure \ref{fig:sub2}) and BeH$_2$ (Figure \ref{fig:sub3}) at bond distances of $r = 1.0$ \AA\, and $r = 3.5$ \AA respectively, were obtained by using the four protocols described above.
The protocols that handle weak symmetric excitations, whether or not they include closure relations, do not achieve chemical accuracy in the studied systems. Protocols handling strong symmetric excitations without closure algebraic relations, converge similarly in both systems, with H$_6$ converging much faster, and requiring fewer BFGS function evaluations in order to reach chemical accuracy (10$^{-3}$ Hartree). In contrast, BeH$_2$ requires about 1500 function evaluations to achieve the same level of convergence.

In the last protocol, where strongly symmetric excitations are considered as starters and incorporate Lie closure relations, we observe that NI-DUCC-VQE allows for a ``full'' convergence of the H$_6$ and BeH$_2$ ground states, with an error of less than 10$^{-10}$ Hartree  relative to the reference FCI energy. The convergence is very rapid, requiring only about 800 function evaluations. These observations suggest that strong symmetric excitations with closure algebraic relations, might approximate strongly correlated states, much better than the NI-DUCC ansatz formed from strongly symmetric excitations missing Lie algebraic closure relations. Moreover, since we simulated BeH$_2$ at a larger bond length of 3.5 \AA, where the ground state is more strongly correlated, we expected to see some difference in the convergence behavior with respect to H$_6$, which was tested near equilibrium at r= 1.0 \AA. However, this was not the case, indicating that the Lie algebraic closure relations help to overcome this challenge.  This demonstrates that incorporating Lie algebraic closure with strong symmetric starters in the NI-DUCC wavefunction systematically provides a solution to the ordering problem. Indeed, several studies  \cite{grimsley2019trotterized, izmaylov2020order, anand2022quantum}  demonstrate that addressing ordering problems can enhance convergence and improve UCC-VQE energy. Similarly, we have shown improvements in the NI-DUCC accuracy thanks to the use of Lie algebraic closure relations.

We then further investigated the importance of  using the ``fourth'' protocol in constructing the NI-DUCC wavefunction at $k=8$. As depicted in Figure \ref{fig:sub2} for H$_6$ and Figure \ref{fig:sub4} for BeH$_2$, we included energy convergence plots not only for $k=8$, but also for $k=3$ to $k=7$. We observed that when $k=3$ and $k=4$ are used, NI-DUCC approaches chemical accuracy (10$^{-3}$ Hartree), while when using $k=5$, $k=6$ and $k=7$, it achieves an accuracy about  $10^{-5}$ Hartree. Although with $k=8$, NI-DUCC-VQE approaches exact FCI solution (with an error  less than 10$^{-11}$ Hartree) after approximately 900 function evaluations without revealing any long plateaus, the differences between various $k$ layers may only matter, if one stops the calculation before a full convergence is reached. This could be the case, for instance, if achieving chemical accuracy is sufficient for a particular molecule. However, if the aim of the computation is to reach full convergence  with the highest level of accuracy, then, using NI-DUCC excitations with sufficient number of layers is advantageous.
Additionally, we have plotted (see Figure. \ref{fig:sub5} for H$_6$ and  Figure. \ref{fig:sub6} for BeH$_2$), the overlap (``fidelity'') of the NI-DUCC-VQE state at each optimization step with the full configuration interaction (FCI) state,  against the number of function evaluations (optimization steps).  Various NI-DUCC tests were performed using different setups: (i) when strong symmetric starters are within closure Lie algebraic relations, we computed NI-DUCC($
k=4$), NI-DUCC($k=6$), NI-DUCC($k=8$);  (ii)  when weak symmetric starters are within closure Lie algebraic relations, we computed NI-DUCC($k=8$). We observed that the NI-DUCC($k=8$) with strong symmetric protocol clearly demonstrates its superiority over ($k=4$ and $k=6$) as well as over weak symmetric ($k=8$) protocol, in predicting more accurate state and energy, as it achieves a fidelity very close to one in less than 200 function evaluations for both molecules. Thus, in the subsequent discussions, we use the NI-DUCC  wavefunction, at $k=8$, employing the fourth protocol outlined earlier, which combines the strong symmetric excitations with a closure Lie algebraic structure.

\subsection{Energy dissociation curves and Energy convergence}
\begin{figure}[!htbp] 
    \begin{subfigure}[t]{0.45\textwidth}
        \centering
    \includegraphics[width=\linewidth]{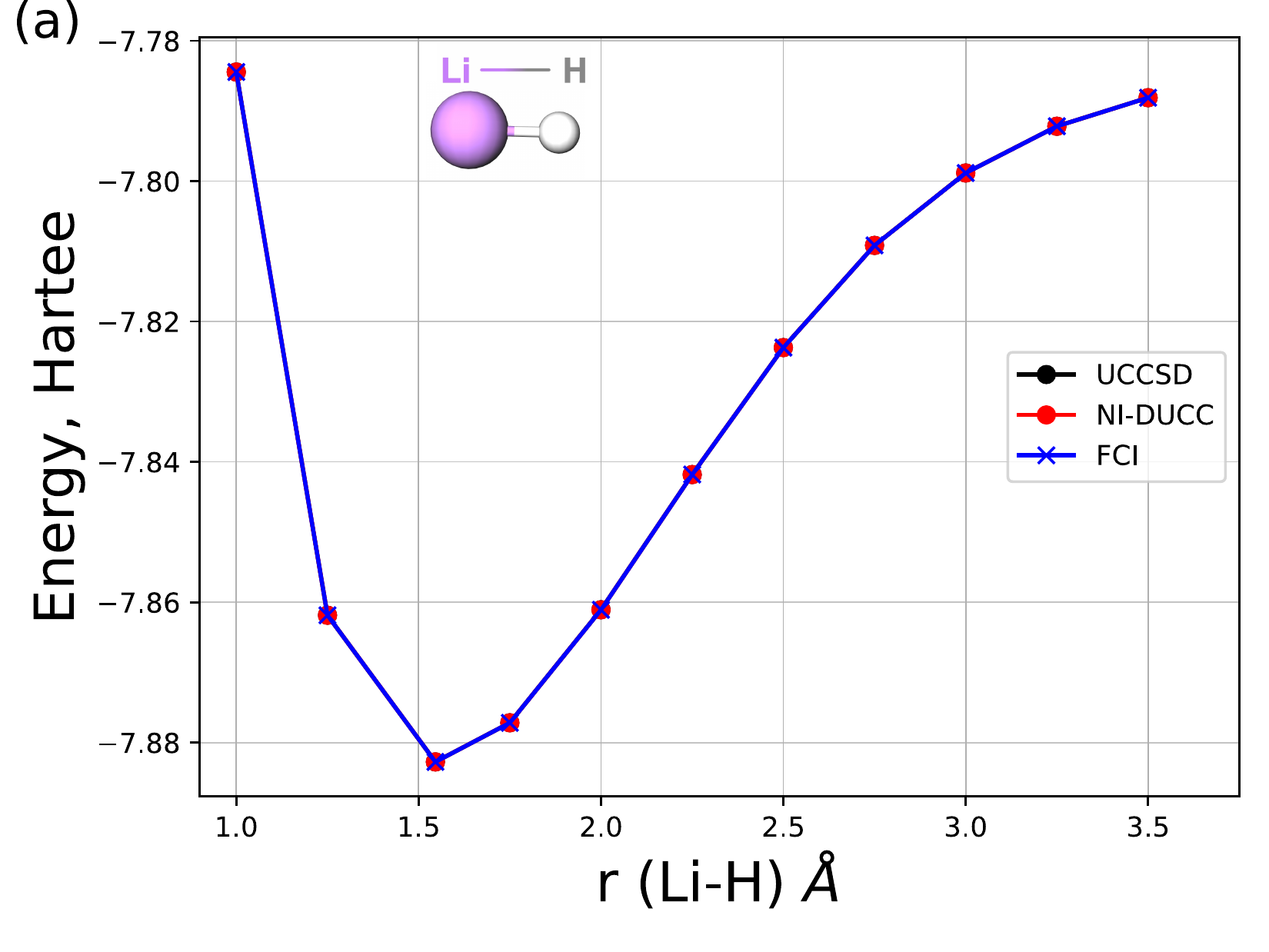}
    \captionlistentry{}
        \label{fig:Lih_diss}
    \end{subfigure}
    \begin{subfigure}[t]{0.45\textwidth}
        \centering
        \includegraphics[width=\linewidth]{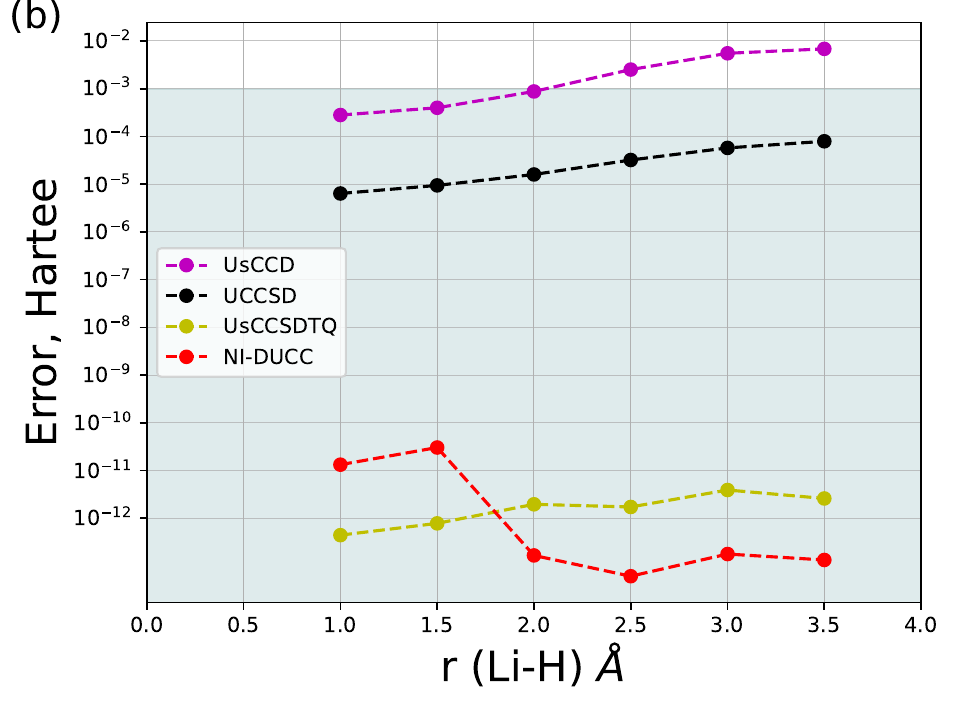}
         \captionlistentry{}
        \label{fig:Lih_comp}
    \end{subfigure}
    \begin{subfigure}[t!]{0.45\textwidth}
        \centering
\includegraphics[width=\linewidth]{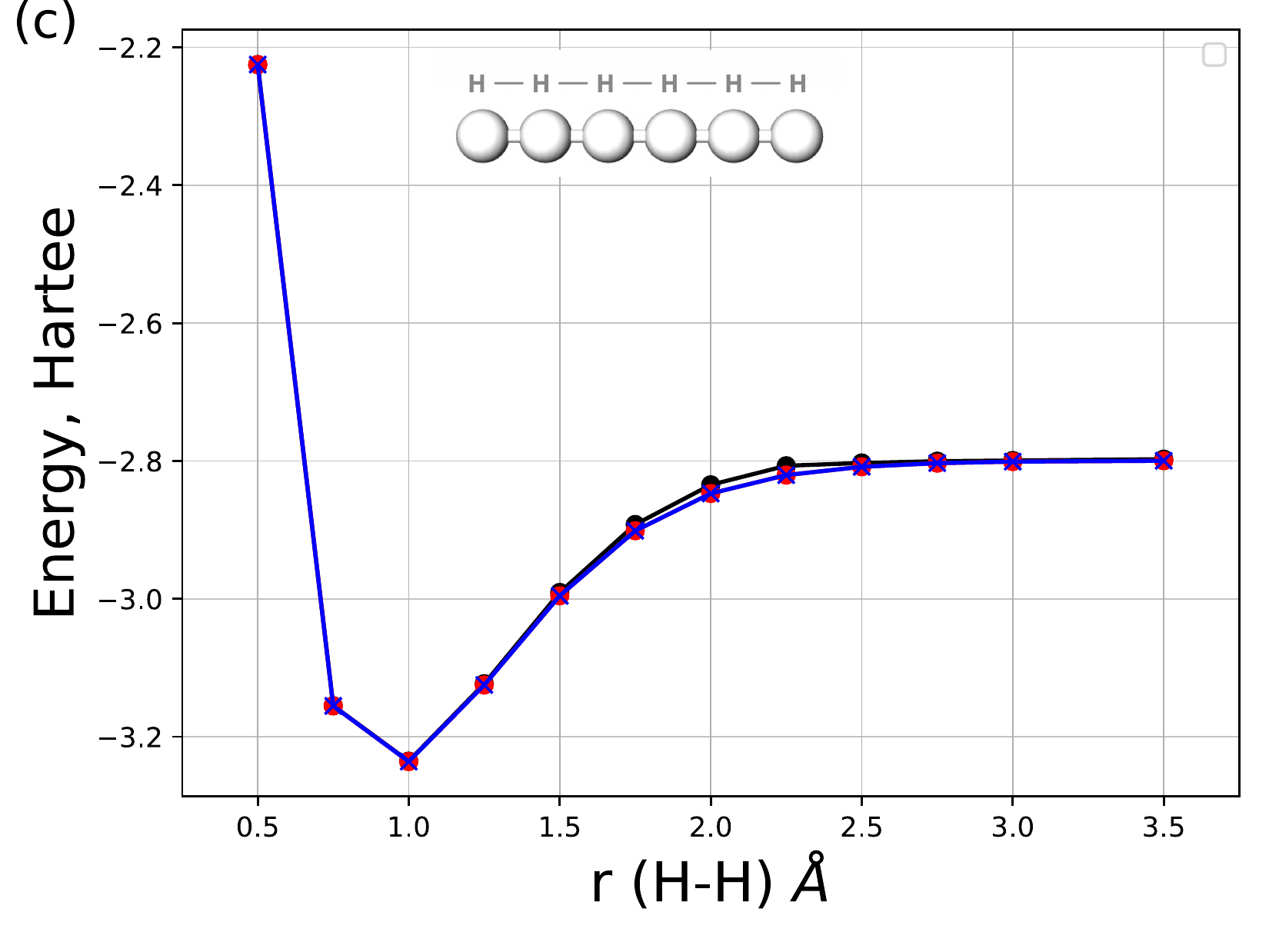}
            \captionlistentry{}
        \label{H6_diss}
    \end{subfigure}
    \begin{subfigure}[t!]{0.45\textwidth}
        \centering
        \includegraphics[width=\linewidth]{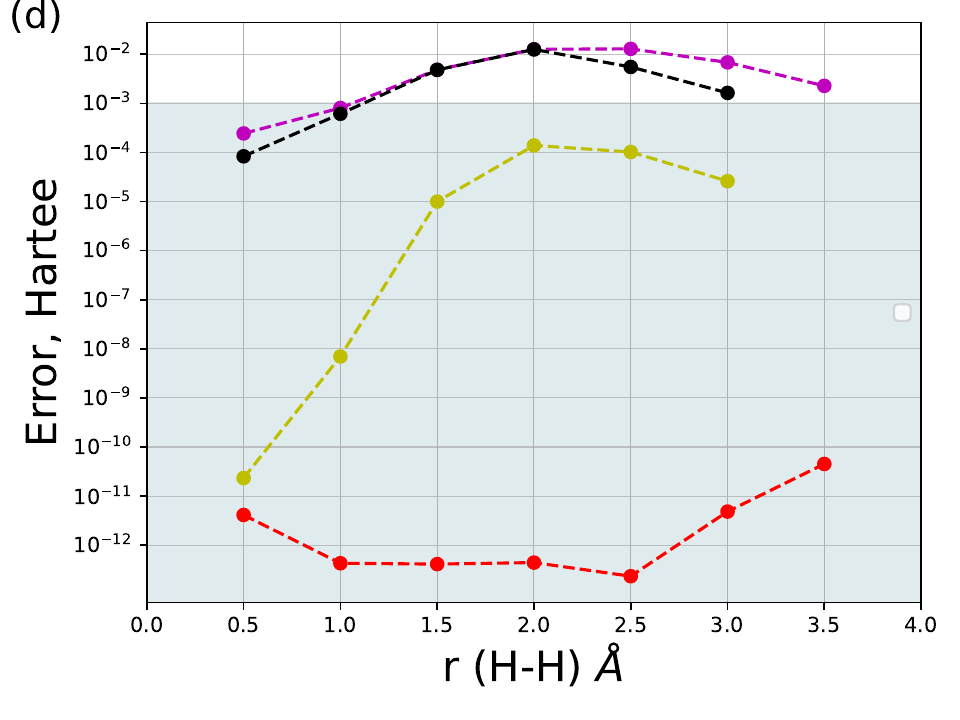}
            \captionlistentry{}
        \label{H6_comp}
    \end{subfigure}
     \begin{subfigure}[t!]{0.45\textwidth}
        \centering
        \includegraphics[width=\linewidth]{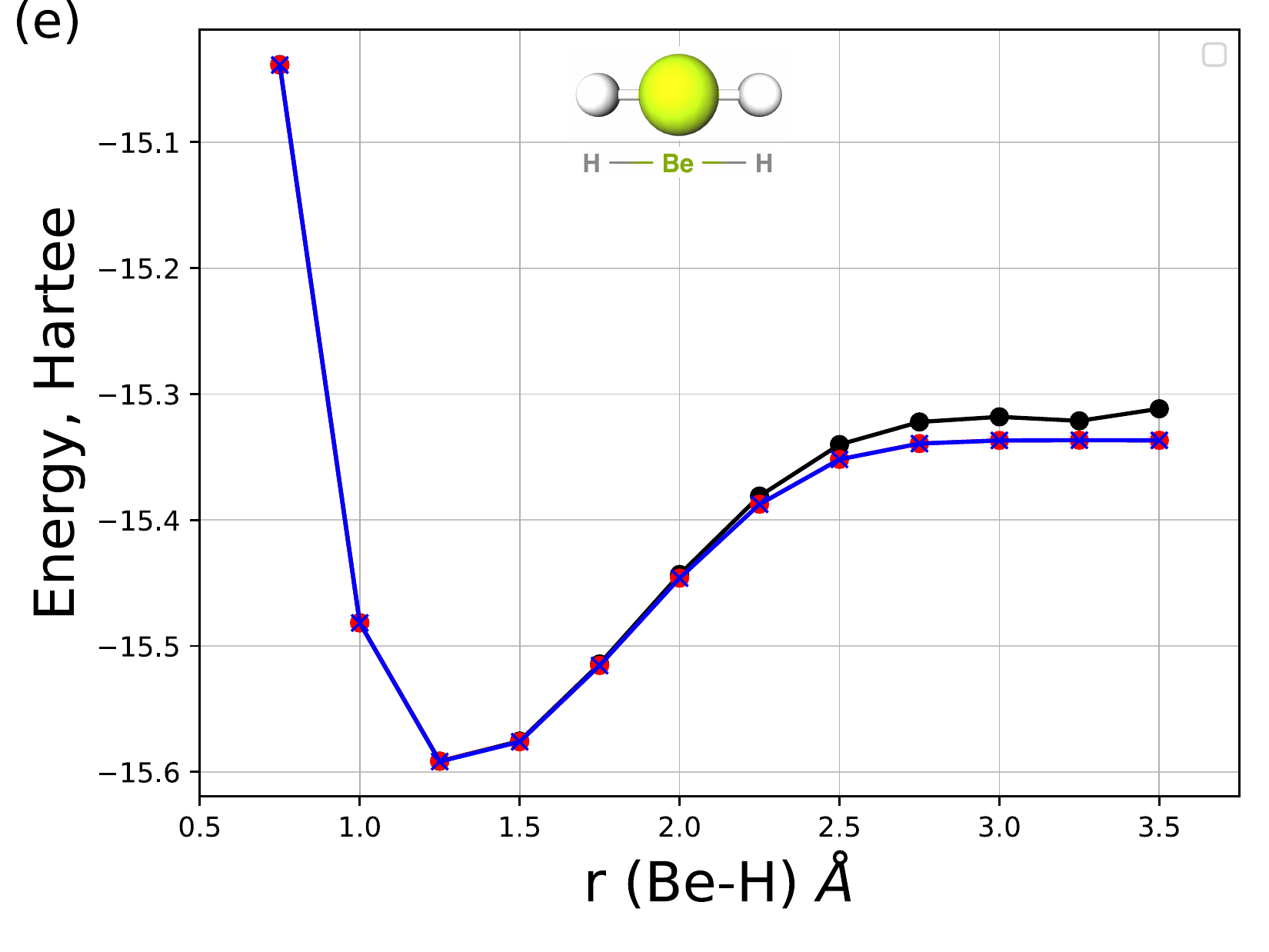}
          \captionlistentry{}
        \label{Beh2_diss}
    \end{subfigure}
    \begin{subfigure}[t!]{0.45\textwidth}
        \centering
        \includegraphics[width=\linewidth]{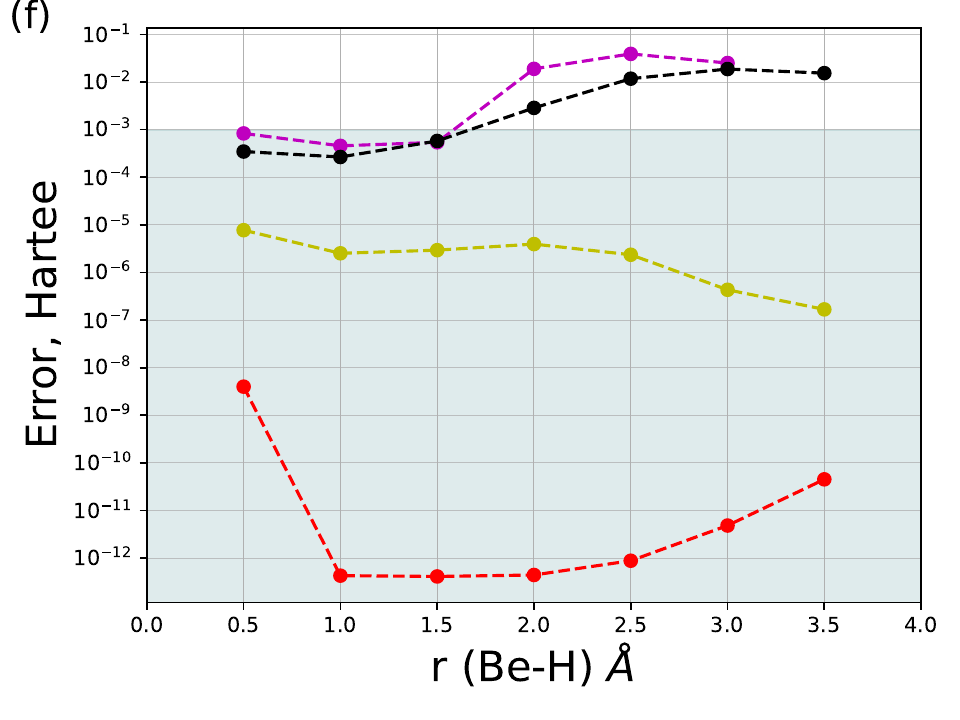}
    \captionlistentry{}
        \label{BeH2_comp}
    \end{subfigure}
    \caption{\RaggedRight \textbf{Dissociation curves performance of 
NI-DUCC-VQE}: for LiH (12 qubits),  H$_6$ (12 qubits) and BeH$_2$ (14 qubits) molecules, in the STO-3G orbital basis set. The energy error is the difference between the obtained energy and FCI solution in (Hartree). The  plots compare the NI-DUCC- VQE ($k=8$) \textbf{(red)}, the UsCCSDTQ-VQE \textbf{(yellow)}, the UCCSD-VQE \textbf{(black)} and the UsCCD-VQE \textbf{(purple)} algorithms. All convergence plots related to UsCCD and UsCCSDTQ are
terminated at an energy threshold equal to $\epsilon = 10^{-8}$ Hartree.}
    \label{fig:main2}
\end{figure}

Figures \ref{fig:Lih_diss}, \ref{H6_diss}, and \ref{Beh2_diss} show the energy dissociation curves and the absolute values for the ground-state energy estimates, for LiH, H$_6$, and BeH$_2$, respectively, obtained by using NI-DUCC-VQE, at $k=8$. The same figures also include the energies of the FCI and untrotterized UCCSD-VQE methods. The three methods yield similar energy estimates, with slight differences observed between UCCSD-VQE and NI-DUCC-VQE at the dissociation bond lengths of H$_6$, and BeH$_2$ molecules.

To make the differences between these methods more evident, figures \ref{fig:Lih_comp}, \ref{H6_comp}, and \ref{BeH2_comp} present the differences  with the exact FCI energy, highlighting the errors of UCCSD and NI-DUCC methods. Additionally, to perform a meaningful comparison, the energy convergence plots comparing these two methods with the unitary selective coupled cluster terminated at quadruple excitations (UsCCSDTQ) \cite{fedorov2022unitary} are provided. UsCCSDTQ-VQE computations were terminated when reaching an  iterative threshold $\epsilon = 10^{-8}$ Hartree. We also included UsCCD-VQE energies using the same threshold. The UsCCD method generates the ``strongly'' double fermionic excitation orbitals, which is used in the first step  while constructing the NI-DUCC-VQE algorithm. The UsCCD-VQE achieves chemical accuracy only over the equilibrium configurations of the three molecules. Alternatively, UCCSD-VQE achieves chemical accuracy over all bond distances for LiH (\ref{fig:Lih_comp}), and at bond distances close to the equilibrium configuration for H$_6$ and BeH$_2$. However, UCCSD-VQE struggles to achieve chemical accuracy for H$_6$ and BeH$_2$ when bond distances exceed 1.5 \AA, where the ground states become more strongly correlated.

Similarly to UsCCSDTQ-VQE, NI-DUCC-VQE achieves chemical accuracy over all investigated bond distances for all three molecules.  
This demonstrates that the two methods can successfully  construct approximate strongly correlated states.  However, the extent to which UsCCSDTQ-VQE deviates from the FCI  solution is ``notable'', especially when the bond lengths start to deviate from equilibrium.  
Over the three molecules, the ansatz constructed by the NI-DUCC-VQE, shows  significant accuracy, with error absolute value less than $10^{-10}$ Hartree, for bond lengths range between 2.0 \AA ~and 3.5 \AA. 
For LiH (Fig. \ref{fig:Lih_comp}), NI-DUCC-VQE outperforms UsCCSDTQ-VQE by approximately one order of magnitude. For H$_6$ (Fig. \ref{H6_comp}), NI-DUCC-VQE outperforms UsCCSDTQ-VQE by approximately seven orders of magnitude. For BeH$_2$ (Fig. \ref{BeH2_comp}), NI-DUCC-VQE outperforms UsCCSDTQ-VQE by an average of four orders of magnitude.

\subsection{Accuracy Limits: NI-DUCC-VQE versus  UsCCSDTQ-VQE and Trotterized sym-UCCSDT-VQE }

The UsCCSDTQ method  \cite{fedorov2022unitary} is designed to recover additional correlation energy by including triple and quadruple excitations. Gradually adding these excitations has been shown to further improve the energy beyond the UCCSD method. However, we observed that, applying UsCCSDTQ-VQE to H$_6$ and BeH$_2$ molecules has some limitations in improving accuracy towards the reference FCI energy solution.

This could be due to the crude approximation used for the screening coefficient computation \cite{fedorov2022unitary}: $|H_{\beta}.t^*_{\beta}| \ge \epsilon$. Such an approximation  is similar to the condition discussed earlier in Section \ref{DUCC-VQE algorithm}, where we first select the dominant fermionic double excitations in NI-DUCC-VQE. As a result, some important excitations appear to be missed during the iteration process, and thus the unitary selective approach, even at the quadruple excitation level, fails to improve the overall accuracy.  

In reference \cite{fedorov2022unitary}, UsCCSDTQ-VQE was tested on the H$_6$ molecule at bond distances $r=0.9$, $1.8$, and $3.6$ \AA{} (see Fig. 2a). It was observed that accuracy challenges worsened as $r$ increased, with UsCCSDTQ deviating further from the FCI solution at distances beyond the equilibrium ($r=0.9$ \AA). In contrast, NI-DUCC was able to reach the FCI solution consistently across all tested distances.

In the same reference \cite{fedorov2022unitary}, UsCCSDTQ-VQE was tested on the BeH$_2$ molecule at $r=3.9$ \AA ( see Fig. 2.C) and it was observed that there is a region where the error remains constant (i.e without improvement in terms of chemical accuracy), despite the addition of more operators to the UsCCSDTQ ansatz. This indicates that some unimportant operators are being added in later iterations ahead of those that would have been more important, to reduce the energy error. This highlights the limitations of the UsCCDTQ approach. Therefore, increasing the iterative threshold ($\epsilon$) further to lower the overall UsCCDTQ-VQE energy appears no necessary.

Similarly, the BeH$_2$ molecule was simulated using the Trotterized sym-UCCSDT method \cite{haidar2023extension} which exploits both spin and point group symmetries. According to our simulations shown in Figure 1.B of reference \cite{haidar2023extension}, the sym-UCCSDT-VQE deviates by approximately 10$^{-3}$ Hartree, and does not show further improvements. We suggested in the article that the sym-UCCSDT approach might miss important higher-order  excitations, i.e beyond triple,  which could be actually contributing significantly to the dynamical correlation, and therefore their absence leads to errors. A similar behavior was observed when sym-UCCSDTQ-VQE was tested on the H$_2$O molecule (see Table 3 in \cite{haidar2023extension}). Indeed these benchmarks demonstrate the challenge of studying highly correlated systems, where many electronic configurations significantly contribute to the wavefunction.

In general, the limitations of triple and quadruple excitations at large bond lengths, could  be potentially addressed by adding higher-order excitations, such as quintuple, sextuple, or septuple. However, this would require substantial effort in the VQE classical optimization, due to the increased number of parameters and to the resulting longer circuit depth, which is undesirable for current NISQ devices. Overall, UsCCSDTQ-VQE and sym-UCCSDT-VQE can construct ansätze to accurately approximate strongly correlated states, but, they do not actually guarantee similar high-level accuracy to the NI-DUCC ansatz. As discussed in Section \ref{numerical}, the Lie algebraic closure with robust starter excitations in the NI-DUCC wavefunction, effectively addresses the ordering problem, in particular at extended bond lengths in H$_6$ and BeH$_2$, thus enhancing the NI-DUCC methodology relative to other ansätze (UsCCSDTQ and sym-UCCSDT) which do not guarantee ordering operators, and thus lead to convergence issues. Overall, NI-DUCC-VQE shows substantial strength in capturing strong correlations at large bond lengths, achieving chemical accuracy within approximately 10$^{-10}$ Hartree of the FCI solution, as illustrated in the figures above.

\section{Computational costs of the NI-DUCC-VQE algorithm} \label{costs}

\subsection{NI-DUCC-VQE versus fixed ansätze:  sym-UCCSDT, UsCCSDTQ, COMPASS and COMPACT}
After establishing NI-DUCC’s high accuracy compared to ansätze with higher-order fermionic excitations (up to quadrupole), we now turn to a detailed comparison. In this subsection, we evaluate NI-DUCC’s parameter efficiency and CNOT gate usage across different molecular systems, contrasting it with UsCCSDTQ and sym-UCCSDT (previously defined in the text), as well as two emerging fixed ansatz methods, COMPASS\cite{mondal2023development} and COMPACT \cite{halder2024noise}. Below, we provide a brief definition of the COMPASS and COMPACT methods:

\begin{enumerate}
    \renewcommand{\labelenumi}{\roman{enumi}.}
    \item The COMPASS ansatz is a dynamically disentangled unitary coupled-cluster protocol that constructs an efficient ansatz using one- and two-body cluster operators alongside selected rank-two scatterers $\sigma$. Its accuracy is tunable through two threshold parameters, (\( \epsilon_1, \epsilon_2\)), that control the inclusion of relevant operators. The ansatz has two versions based on the choice of the $\sigma$ operator in the selection bath: (a) the opposite-spin (OP) sector of $\sigma$ and (b) the partially paired (PP) sector, termed COMPASS-OP and COMPASS-PP, respectively (see Section III in \cite{mondal2023development}). Both versions can utilize parallel quantum architectures with energy-based sorting and commutativity prescreening to enhance efficiency and minimize gate depth.
    \item COMPACT ansatz is a threshold-based, noise-robust strategy built on \textit{ab-initio} many-body perturbation theory, and does not require any pre-circuit measurements. Its accuracy and quantum complexity are tunable by setting a perturbative order and threshold parameters, (\( \epsilon_1, \epsilon_2, \epsilon_3 \)), which filter for the most significant excitations. This structure decomposes high-rank excitations into low-rank operators, which helps in optimizing computational resources and in minimizing gate depth.
\end{enumerate}

We tested the efficacy and accuracy of the NI-DUCC method on three prototypical systems: LiH (at \( r = 1.0 \, \text{\AA} \)), BeH\(_2\) (at \( r = 2.0 \, \text{\AA} \)), and BH (at \( r = 2.0 \, \text{\AA} \)). These molecules and bond lengths were selected to enable a fair comparison with the COMPASS\cite{mondal2023development} and COMPACT \cite{halder2024noise} ansätze by drawing on data from their respective Figure 2.

As shown in Table 1, UsCCSDTQ achieves high accuracy (\( 10^{-7} \)) with only 61 parameters, but at the cost of a very high CNOT gate count (26,608). In contrast, the sym-UCCSDT ansatz maintains accuracy with only 58 parameters and significantly fewer CNOT gates (14,304). Notably, COMPACT(5,5,4) requires just 44 parameters and achieves a much lower gate count (3,580) compared to sym-UCCSDT, underscoring its efficiency.
The NI-DUCC method exhibits a progressive improvement in chemical accuracy with increasing layer depth \( k \): from \( 10^{-5} \) at \( k = 6 \), to \( 10^{-6} \) at \( k = 7 \), and to \( 10^{-12} \) at \( k = 8 \). Additionally, NI-DUCC achieves a significant reduction in gate count approximately by factor 3 compared to COMPACT(5,5,4) although the parameter count increases with additional layers.

For BeH\(_2\) (See Table 2), the sym-UCCSDT ansatz achieves a chemical accuracy of only \(10^{-4}\) relative to FCI, while UsCCSDTQ improves this by two orders of magnitude, highlighting the impact of quadruple excitations. However, higher-order excitations also lead to a significant increase in the CNOT gate count as noticed in Table 2.
In contrast, COMPASS-OP(5,5,4) and COMPASS-PP(5,5,4), which each  uses up to two-body cluster operators, require only 54 parameters and reduce the CNOT gate count by approximately an order of magnitude compared to sym-UCCSDT.
The NI-DUCC ansatz, at \(k=6\) and \(k=8\), improves accuracy from \(10^{-5}\) to \(10^{-7}\) with a minimal number of CNOT gates. At \(k=8\), it reaches an accuracy of \(10^{-12}\) with only 1,880 CNOT gates, in contrast to COMPASS-(5,5,4), which requires 2,592 gates to achieve \(10^{-5}\), where 7 orders of magnitudes differ. However, NI-DUCC at this level necessitates roughly five times more parameters than COMPASS-(5,5,4).

The results for BH (Table 3) also demonstrate the NI-DUCC ansatz's efficiency in minimizing CNOT gate counts compared to COMPASS and COMPACT. For an accuracy of $10^{-4}$, COMPASS-(5,7)-OP requires 2,736 CNOT gates, whereas NI-DUCC at $k=4$ achieves the same accuracy with only 680 CNOTs. Increasing NI-DUCC layers to $k=5$ improves accuracy by an order of magnitude with just 850 CNOTs.
For $10^{-6}$ Hartree, COMPACT(3,3,4) requires 5,048 CNOTs and 59 parameters, while NI-DUCC at $k=6$ achieves this accuracy with 1,020 CNOTs and 204 parameters. NI-DUCC further enhances accuracy to $10^{-12}$ with only 1,360 CNOTs, considerably fewer than COMPASS or COMPACT, though it demands a higher parameter count.
These results highlight NI-DUCC's strong performance in reducing CNOT gates while maintaining a high level of chemical accuracy in the simulated molecules, outperforming the fixed ansatz approaches. However, they also indicate a need to further optimize the parameter count, which could potentially be improved with future techniques.

\begin{table}[h!]
\centering
\arrayrulecolor{black}  
\renewcommand{\arraystretch}{1.7} 
\hspace*{-1.3cm} 
\begin{tabular}{|c|c|c|c|c|c|c|}
\hline
\multicolumn{1}{|c|}{$\bm{\text{Molecule}}$} &$\bm{\text{UsCCSDTQ}}$& $\bm{\text{sym-UCCSDT}}$  & $\bm{\text{COMPACT(5,5,4)}}$ & $\bm{\text{NI-DUCC($6$)}}$ & $\bm{\text{NI-DUCC($7$)}}$ & $\bm{\text{NI-DUCC($8$)}}$ \\
\hline
LiH  & (61) (26608) $ 10^{-7}$ &  (58)(14304) $10^{-7}$&(44) (3580) $10^{-7}$      & (138) (690) 
 $10^{-5}$  & (161) (805) $10^{-6}$ & (184) (920) $10^{-12}$ \\
 \hline
\end{tabular}
\caption{\RaggedRight Comparison of $\bm{\text{sym-UCCSDT}}$, $\bm{\text{UsCCSDTQ}}$, $\bm{\text{COMPACT(5,5,4)}}$ and $\bm{\text{NI-DUCC}}$ at layers $k = 6, 7, 8$ for BeH$_2$ geometry at bond length $r_{\text{Li-H}}$ = 1.0 \AA \,in the STO-3G basis. Estimations provided in each box include parameter count, CNOT count, and energy error relative to FCI (Hartee) (Same for Table II and Table III). Data for $\bm{\text{UCCSDT}}$, $\bm{\text{sym-UCCSDT}}$ and $\bm{\text{COMPACT(5,5,4)}}$ are extracted from \cite{halder2024noise}, where COMPACT uses thresholds $\log(\varepsilon_1)$, $\log(\varepsilon_2$) and $\log(\varepsilon_3$)  to select significant spin-complementary excitations in the ansatz. Our UsCCSDTQ simulations are terminated at a threshold of $\epsilon = 10^{-8}$ Hartree. CNOT count scaling details are available in Supplementary Table 1 of Supplementary Materials~\cite{thisSuppl} for all ansätze.}
\label{tab:countSF1}
\end{table}
\begin{table}[h]
\centering
\arrayrulecolor{black}  
\renewcommand{\arraystretch}{1.7} 
\hspace*{-1.5cm} 
\begin{tabular}{|c|c|c|c|c|c|c|c|}
\hline
\multicolumn{1}{|c|}{$\bm{\text{Molecule}}$}  & $\bm{\text{UsCCSDTQ}}$ &   $\bm{\text{sym-UCCSDT}}$ &$\bm{\text{COMPASS(5,7)-OP}}$&$\bm{\text{COMPASS(5,7)-PP}}$ & $\bm{\text{NI-DUCC($6$)}}$ & $\bm{\text{NI-DUCC($7$)}}$ & $\bm{\text{NI-DUCC($8$)}}$ \\
\hline
BeH$_2$ & (84) (44840) $10^{-6}$ & (146)(21696) $10^{-4}$  &(54) (2592) $10^{-5}$ & (54) (2592) $10^{-5}$  & (282) (1410) $10^{-5}$ & (329) (1645) $10^{-6}$ & (376) (1880) $10^{-12}$ \\

\hline
\end{tabular}
\caption{\RaggedRight  Comparison of $\bm{\text{UsCCSDTQ}}$, $\bm{\text{COMPASS(5,7)-OP}}$, $\bm{\text{COMPASS(5,7)-PP}}$, and $\bm{\text{NI-DUCC}}$ at layers $k = 6, 7, 8$ for BeH$_2$ geometry at bond length $r_{\text{Be-H}}$ = 2.0 \AA \,in the STO-3G basis.  Data for $\bm{\text{COMPASS(5,7)-OP}}$ and $\bm{\text{COMPASS(5,7)-PP}}$, are extracted from \cite{mondal2023development}, where COMPASS uses thresholds $\log(\varepsilon_1)$ and $\log(\varepsilon_2$) to select significant cluster amplitudes and scatterers in the ansatz. }
\label{tab:countSF2}
\end{table}
\begin{table}[h]
\centering
\arrayrulecolor{black}  
\renewcommand{\arraystretch}{1.7} 
\hspace*{-1.3cm} 
\begin{tabular}{|c|c|c|c|c|c|c|c|}
\hline
\multicolumn{1}{|c|}{$\bm{\text{Molecule}}$}  & $\bm{\text{COMPASS(5,7)-OP}}$&$\bm{\text{COMPACT(3,3,4)}}$ & $\bm{\text{NI-DUCC($4$)}}$ & $\bm{\text{NI-DUCC($5$)}}$ & $\bm{\text{NI-DUCC($6$)}}$&$\bm{\text{NI-DUCC($7$)}}$&$\bm{\text{NI-DUCC($8$)}}$ \\
\hline
BH & (57) (2736) $10^{-4}$  &  (59) (5048) $10^{-6}$  &(136) (680) $10^{-4}$      & (170) (850) 
 $10^{-5}$  & (204) (1020) $10^{-6}$ & (238) (1190)$10^{-7}$ &(272) (1360) $10^{-12}$ \\
 \hline
\end{tabular}
\caption{\RaggedRight Comparison of $\bm{\text{COMPASS(5,7)-OP}}$ $\bm{\text{COMPACT(3,3,4)}}$ and $\bm{\text{NI-DUCC}}$ at layers $k = 4, 5, 6,7,8$ for BH geometry at bond length $r_{\text{B-H}}$ = 2.0 \AA \,in the STO-3G basis. Data for $\bm{\text{COMPASS(5,7)-OP}}$ and $\bm{\text{COMPACT(3,3,4)}}$  are extracted from figure 2 of \cite{mondal2023development} and \cite{halder2024noise} respectively.}
\label{tab:countSF3}
\end{table}

\subsection{NI-DUCC-VQE versus ADAPT-VQE}
\begin{figure}[!htbp]
    \centering
    \begin{subfigure}[b]{0.49
\textwidth}
        \centering
        \includegraphics[width=\textwidth]{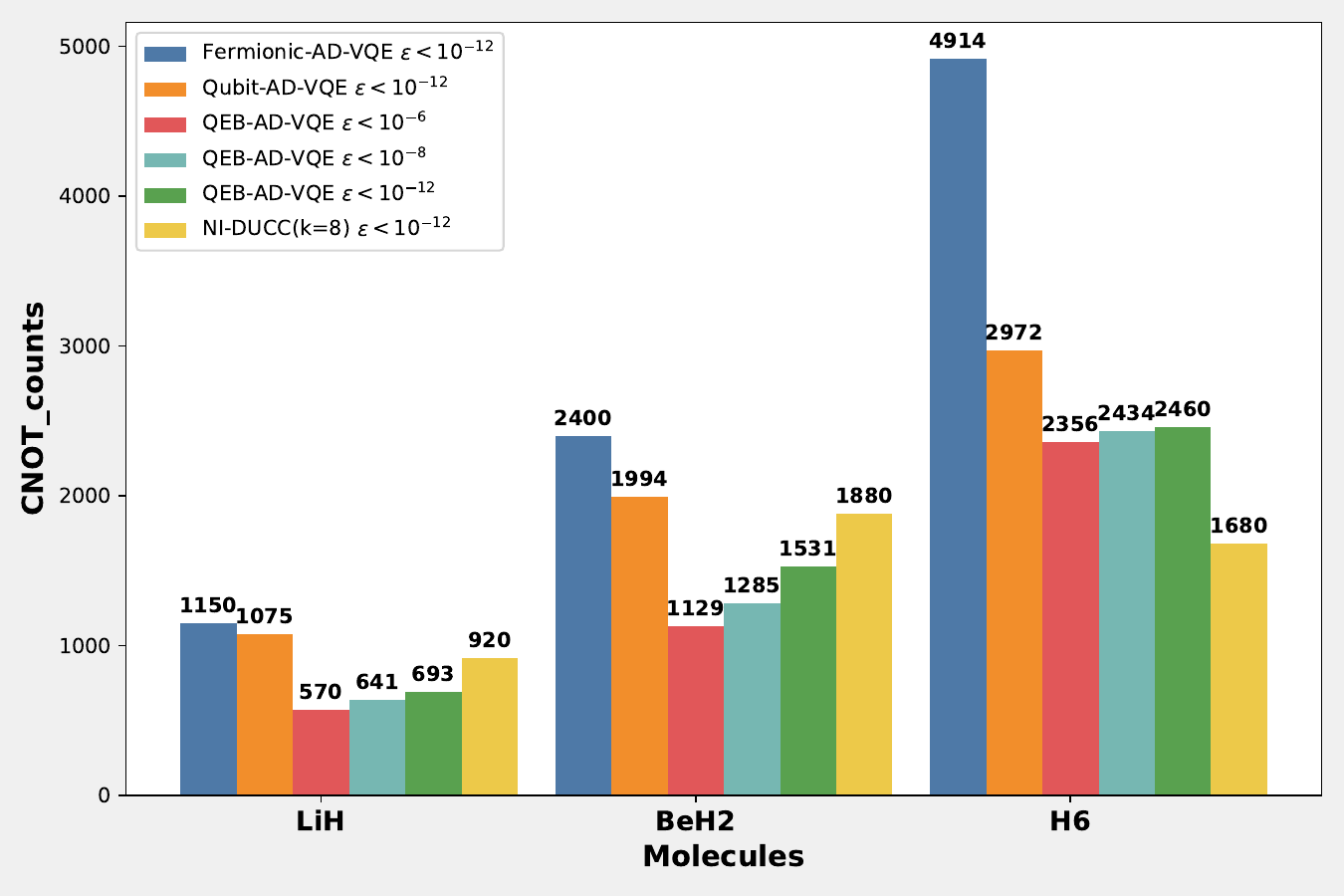}
        \caption{CNOT counts}
        \vspace{0.5cm}
        \label{fig:left1}
    \end{subfigure}
     \begin{subfigure}[b]{0.49\textwidth}
        \centering
        \includegraphics[width=\textwidth]{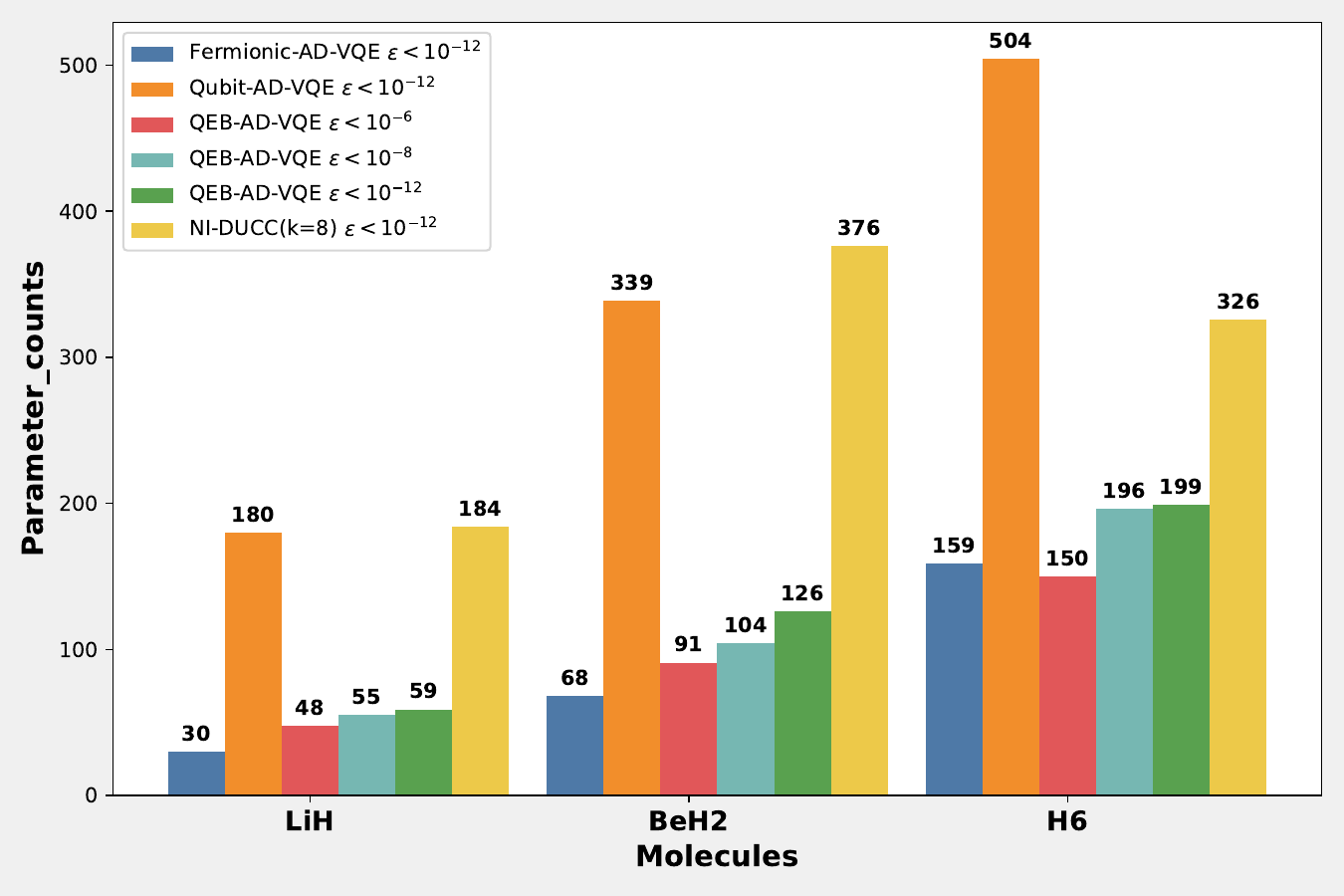}
        \caption{Parameter counts}
        \vspace{0.5cm}
        \label{fig:right1}
    \end{subfigure}
    \vspace{1.0cm}
    \begin{subfigure}[b]{0.49\textwidth}
        \centering
        \includegraphics[width=\textwidth]{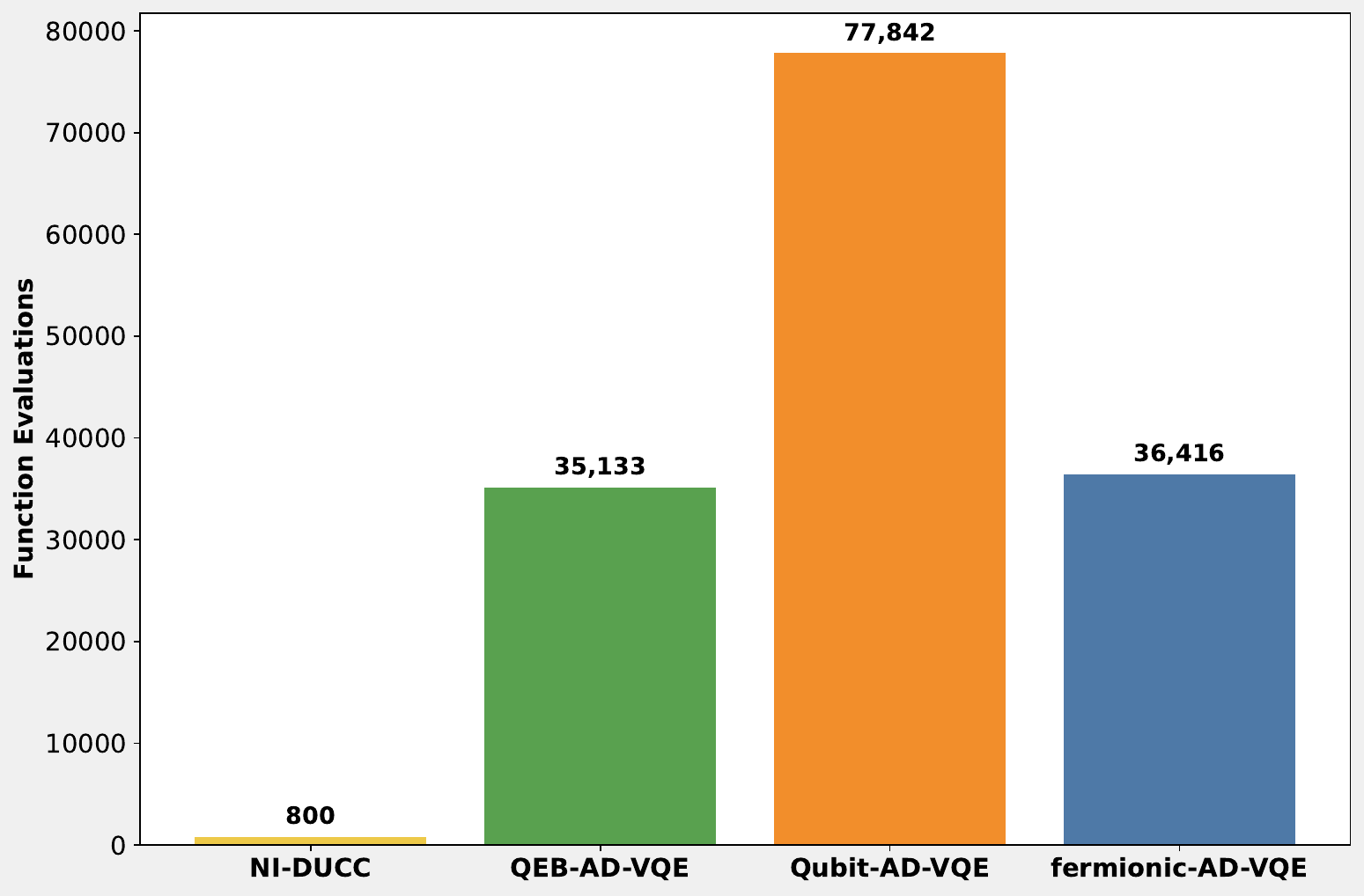}
        \caption{H$_6$}
        \label{fig:left2}
    \end{subfigure}
     \begin{subfigure}[b]{0.49
\textwidth}
        \centering
        \includegraphics[width=\textwidth]{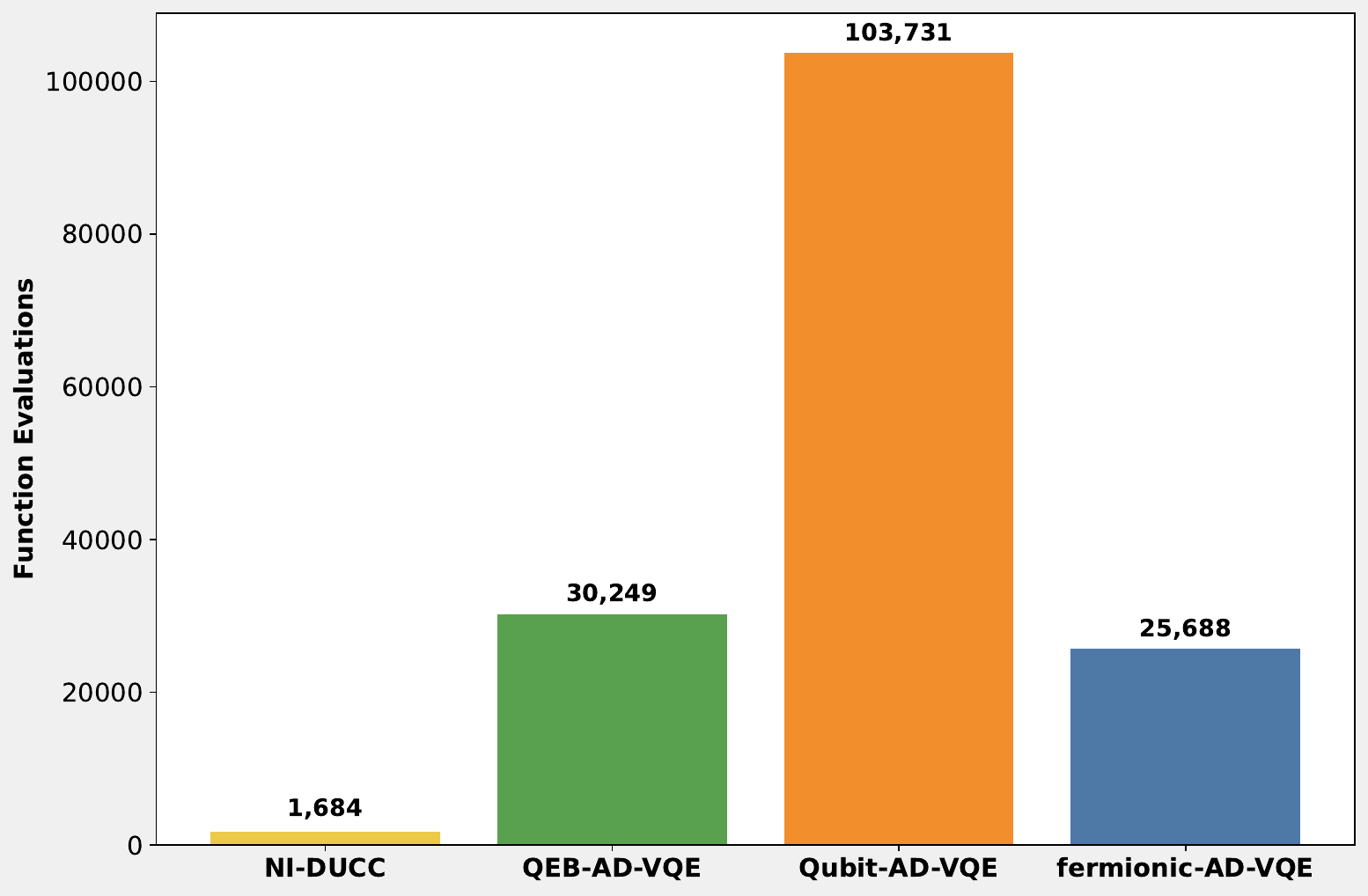}
        \caption{BeH$_2$}
        \label{fig:right2}
    \end{subfigure}    
    \caption{\RaggedRight \textbf{Resource comparison of the QEB-ADAPT-VQE, the fermionic-ADAPT-VQE, the qubit-ADAPT-VQE and NI-DUCC-VQE}.  The tested molecules (STO-3G) are LiH  ($r_{\text{Li-H}}$ = 1.546 \AA), BeH$_2$ ($r_{\text{Be-H}}$ = 1.316 \AA) , and H$_6$ ( $r_{\text{H-H}}$  = 1.5 \AA,).  
    \textbf{(a)} CNOT counts, \textbf{(b)}  Parameter counts, \textbf{(c,d)}
 function evaluation  for H$_6$ (c) and BeH$_2$ (d), respectively. Note that the number of operators are equal to the number of parameters in each algorithm. Also, the function evaluations are the number of optimization steps in the BFGS optimizer.}
    \label{fig:main}
\end{figure}

 Given that ADAPT-VQE algorithms have demonstrated significant improvements in term of accuracy, as well as  a reduction in computational costs (see Table S1 in the Supplementary Materials~\cite{thisSuppl}), in this subsection we will focus on comparing the computational costs of NI-DUCC-VQE with fermionic-ADAPT-VQE  \cite{grimsley2019adaptive}, qubit-ADAPT-VQE  \cite{tang2021qubit}, and QEB-ADAPT-VQE  \cite{yordanov2021qubit}. To ensure a fair comparison, we use data for LiH, H$_6$, and BeH$_2$ at bond distances of 1.546 \AA, 1.5 \AA, and 1.316 \AA, respectively (as shown in Figure 6 of reference  \cite{yordanov2021qubit}). The numerical results were extracted from the link provided in \cite{JordanovSJ}. Based on this data, we perform NI-DUCC calculations on the same molecules bond distances, and their estimate computational resources (see Figure \ref{fig:main}).
We set $k=8$ for NI-DUCC-VQE to achieve a similar accuracy to the ADAPT-VQE algorithms, with an energy threshold of $10^{-12}$ Hartree. Moreover, since NI-DUCC is a fixed ansatz, requiring only one iteration and one fixed circuit execution, we do not compare it with ADAPT-VQE in terms of the number of iterations (see Section H in reference  \cite{yordanov2021qubit}, where the authors discuss the total number of iterations required by each ADAPT-VQE algorithm).

In Figure \ref{fig:left1}, at the $10^{-12}$ Hartree level of accuracy, where larger circuits are expected, the QEB-VQE-ADAPT method systematically outperforms both fermionic- and qubit-ADAPT-VQE in terms of CNOT efficiency. This superiority stems from the type of pool excitations used in the ansatz. According to   \cite{yordanov2021qubit}, QEB-ADAPT-VQE employs efficient qubit excitation circuits, that are more effective than the fermionic excitations used in fermionic ADAPT-VQE. Moreover, qubit-ADAPT-VQE fails to surpass QEB-ADAPT-VQE energies at high levels of accuracy, simply because qubit evolutions enable local circuit optimizations, while the more rudimentary Pauli string evolutions used by qubit-ADAPT-VQE do not. Therefore it would require more iterations to approach the exact energy.
The NI-DUCC CNOT counts exhibit similar results to QEB-ADAPT. This is due to the fact that the NI-DUCC ansatz elements consist in evolutions of $XY$-Pauli strings of length
$p$ with an odd $Y$ Pauli terms, leading to CNOT gates scaling as $2p-2$ per Pauli string \cite{tang2021qubit}. This indicates that the NI-DUCC excitations ansatz achieves an efficient scaling in the usage of CNOT gates.

In terms of parameter counts, while NI-DUCC-VQE and QEB-ADAPT-VQE require significantly fewer CNOTs compared to fermionic-ADAPT-VQE, the QEB-ADAPT-VQE requires up to twice as many variational parameters, and the NI-DUCC-VQE requires up to three times as many (see Fig. \ref{fig:right1}). This difference arises because NI-DUCC-VQE and QEB-ADAPT-VQE, assign one parameter to each qubit excitation (one Pauli string)  and qubit excitation  (single and double) evolutions in their ansatz, respectively, whereas fermionic-ADAPT-VQE assigns one parameter to a pair of spin-complement fermionic excitation evolutions. These spin-complement pairs can enforce parity conservation, which is not achieved by qubit excitation evolutions. Additionally,  NI-DUCC requires more parameters due to the dynamic increase in the number of layers $k$ of the ansatz (see Eq. \ref{disentangled1}), even though the pool generators are complete and originate from the symmetric starter excitations, which  conserves the parity symmetry property in NI-DUCC ansatz. The reader can refer to Table S1 in Supplementary Materials~\cite{thisSuppl} which compares  the scales of NI-DUCC, UsCCSDTQ, and ADAPT-VQE versions in terms of CNOT and parameter counts.

We also computed the number of function evaluations required by each of the studied algorithms, as shown in Figures \ref{fig:left2} for H$_6$ and \ref{fig:right2} for BeH$_2$. As noted in the computational procedure Section, all algorithms used the BFGS optimizer, with analytically calculated energy gradient vectors. In both benchmarks, we observe that the NI-DUCC-VQE algorithm requires only 800 function evaluations for H$_6$, and 1684 for BeH$_2$, to achieve an accuracy below 10$^{-12}$
  Hartree. In contrast, the ADAPT-VQE algorithms need at least 25 $\times 10^{3}$ to 30 $\times 10^{3}$
  function evaluations to reach the same level of accuracy.
This highlights the advantage of NI-DUCC-VQE in achieving full convergence, with a relatively small number of optimization steps, i.e. function evaluations. This efficiency is attributed to the combination of strongly symmetric correlated orbitals in the initial guess, with the closure algebraic relations, which accelerates the computations, and avoids false local minima, immune to barren plateaus, as observed in both molecules (see Figures \ref{fig:sub2} and \ref{fig:sub3} in Section \ref{numerical}). 

Overall, when considering algorithmic costs, it is crucial to assess the CNOT gate counts, the parameter counts and function evaluations. While fewer CNOT gates reduce the circuit depth and enhance robustness against noise, a higher parameter count can improve convergence rates, as observed in our simulations. Specifically, NI-DUCC-VQE has demonstrated rapid convergence, and achieving FCI solutions, with significantly fewer function evaluations compared to ADAPT-VQE methods. Therefore we can say while NI-DUCC may require more parameters, its efficient CNOT usage and rapid convergence demonstrate that this trade-off can lead to superior performance in practical quantum computing hardware implementation scenarios.

\subsection{Challenges in MCP generation as the number of qubits increases}
\label{challenges}
The main cost in constructing the NI-DUCC ansatz, lies in the precomputation step that involves building the symmetry-preserving MCP. 
The MCP consists of a set of generators that scale as $\mathcal{O} (n)$ with $n$ qubits, and can form a Lie algebraic closure set, as previously explained in the methodology. 
The definitions, properties, completeness conditions of MCP along with its symmetry-related conditions, are described in section 4 of the Supplementary Materials \cite{thisSuppl} (note that comprehensive information about MCP generation can be also found in the reference  \cite{shkolnikov2023avoiding}).
The main steps involved in constructing the MCP and their associated computational costs on a classical computer, are as follows:
\begin{enumerate}
    \item \textbf{Generate the complete pool, the FullGroup, and the FullSet} (see Figure S1 in the supplementary materials \cite{thisSuppl}). The complete pool scales as $2^{2(n-1)}$, which is generated by constructing the  product group of the set $\mathcal{S}= \left\{Z_1,\ldots,Z_{n-2},Y_1,\ldots,Y_{n-1},Z_{n-1}Y_n \right\}$. The FullGroup and FullSet, are both essential computations used to validate the completeness of the pool. Each of them  contains elements from the complete pool that are constrained by symmetries. Although the FullGroup and the FullSet can scale less due to the added symmetries, they still exhibit exponential growth with $n$. For example, the FullGroup scales numerically as $2^{2(n-3)}$.
    
    \item \textbf{Generate an initial set with starters that are symmetry-preserving by incorporating particle number, spin, parity and spatial symmetry}. These starters need to involve double-excitations, as they act on top of the Hartree-Fock state. Numerically, one might encounter a pool, larger than $2n-2$, due to the number of initial starters, but the scaling remains linear.
    
    \item \textbf{Add $m$ randomly selected  Pauli strings from the FullSet to the initial set of starters, where} $m\in  \mathbb{N}^*$. 
    \item \textbf{Verify the completeness conditions of the newly generated set.} This involves verifying two conditions: first, ensuring the size of the product group generated by the new set matches that of the FullGroup, and second, confirming the inseparability condition, where the set cannot be split into two mutually commuting sets. If these conditions are not met, repeat step 3.
\end{enumerate}

The time and memory required to generate and process the product group and Lie algebra through string multiplications grow exponentially with the number of qubits. We also refer readers to step d in Figure \ref{figure1} for visual clarification and to Section 4 of the supplementary material \cite{thisSuppl} for a more detailed explanation.

In terms of coding, the Python code for generating the MCP can be found in reference \cite{VladShkolnikov}. However, as mentioned in the computational details (See Section \ref{computational details}), we transformed the Python code into a C++ program. Indeed, comparing the performance of Python and C++ implementations, there are several critical factors come into play in the context of High-Performance Computing (HPC): execution speed, memory consumption, and ease of optimization (see further explanations of these factors in section 5 of the Supplementary Materials \cite{thisSuppl}).
In our tests, storing  for example the FullGroup using python as a python set, is memory inefficient when ones tries to scale up the number of qubits. To overcome this issue, we designed a specific C++ data structure set, that minimizes memory consumption to the detriment of searching operations, in order to be able to hold larger sets and to  increase the number of qubits. This grants the ability to run 14 qubits systems encompassing 7 electrons, on a local laptop, without suffering from memory issues. In practice,  the C++ approach only requires 95 MiB instead of the 14 GiB required by the python code. Our C++ subroutine, is fully optimized, which enables to generate MCPs for up to 20 qubits while insuring sufficient large amount of memory. However, with the present implementation, the time required to perform product groups computations (i.e between 20 and beyond) remains very high. 
Table \ref{tab:countSF4} provides an example of the cost associated with the lists generation for the FullSet, the FullGroup, and the full symmetric starters, for constructing the MCP. It is evident that these costs scale exponentially with $n$,  in both computer memory and computational time (see Figures S3 (a) and S3 (b) of the Supplementary Materials \cite{thisSuppl}): 

\begin{enumerate}
\item \textbf{Memory Cost}. For example, if we aim to generate an MCP associated to 18 qubits, this would require to store a large number of excitations: in the full starters (3.34 $\times 10^7$), FullSet (1.07 $ \times 10^9$) and FullGroup (2.15 $ \times 10^9$), each represented by a list of Pauli strings. Storing them would require in total 24.3 GiB of memory, using the present C++ subroutine. As the number of qubits increases to 22, storing all possible excitations requires an exponentially growing amount of memory equivalent to 6 TB.
\item \textbf{Time Cost.} For 18 qubits, the time required to generate and process the  product group operations takes approximately 6 hours of CPU time,  and it scales exponentially with an increasing number of qubits.
\end{enumerate}

\begin{table}[h]
\centering
\arrayrulecolor{black}  
\begin{tabular}{|c|c|c|c|c|}
\hline
\multicolumn{1}{|c|}{$\bm{\text{(qubits,electrons)}}$} & $\bm{\text{Minimal Pool (MCP)}}$ & $\bm{\text{Number of starters}}$ & $\bm{\text{Number of FullSet}}$ & $\bm{\text{Number of FullGroup}}$ \\
\hline
(8,4)    & 14 & 320          & 992            & 2048           \\
(10,5)   & 18 & $2.82 \times 10^3$ & $1.63 \times 10^4$  & $3.28 \times 10^4$  \\
(12,6)  & 22 & $2.61 \times 10^4$ & $2.62 \times 10^5$  & $5.24 \times 10^5$  \\
(14,7)   & 26 & $1.88 \times 10^5$ & $4.19 \times 10^6$  & $8.39 \times 10^6$  \\
(16,8)   & 30 & $1.38 \times 10^6$ & $6.71 \times 10^7$  & $1.34 \times 10^8$  \\
(18,9)   & 34 & $3.43 \times 10^7$ & $1.07 \times 10^9$  & $2.15 \times 10^9$  \\
(20,10)  & 38 & $1.01 \times 10^9$ & $1.72 \times 10^{10}$ & $3.44 \times 10^{10}$ \\
(22,11)  & 42 & $2.56 \times 10^{10}$ & $2.74 \times 10^{11}$ & $5.50 \times 10^{11}$ \\
\hline
\end{tabular}
\caption{\RaggedRight The estimated total number of starters,  size of the FullSet and the FullGroup required to generate a minimal complete pool, per system, for a given number of qubits and electrons.  Calculations are performed on a standard computer machine: \textit{2 CPUs Intel(R) Xeon(R) Silver 4116 CPU @ 2.10GHz  12 Cores  + 131 GiB. OS: CentOS Linux 7}. Note that the starters here exhibit good symmetry but are not necessarily strongly symmetric. It includes all weak and strong excitations.}
\label{tab:countSF4}
\end{table}
Despite the discussed C++ optimization of the MCP generation, which allowed to specifically accelerate the construction of the product group for qubits ranges between 14 and 20, the MCP generation still imposes a significant computational load on classical computers. Presently, such an implementation becomes impractical as the number of qubits exceeds 20.

The MCP generation corresponds actually to the final step of the NI-DUCC-VQE algorithm (see Section 2.\ref{DUCC-VQE algorithm}). Overall, the other steps are not particularly costly, as we only generate the starters of double excitations using sparsity conditions, and then impose symmetry preservation, on the qubit excitations of these starters. These steps, along with a symmetry-preserving MCP generation, rely solely on classical computation. Therefore, the NI-DUCC ansatz is dynamically constructed with multiple layers (see Eq. \ref{disentangled1}), without any ``pre-circuit'' measurements from the quantum computer. Indeed, once the computational load for MCP generation beyond 20 qubits is managed, NI-DUCC’s fixed, non-iterative ansatz will keep costs low, requiring only standard VQE-style measurements. Unlike ADAPT-VQE, which incurs high costs from iterative gradient measurements, NI-DUCC leverages symmetry constraints with Lie algebraic conditions, which enables faster convergence and efficient scaling.
Currently, the time cost for DI-DUCC at 18 qubits reflects our specific simulation setup and appears costly. However, direct comparisons across algorithms remain challenging due to differences in the followings: resource requirements, parameter scaling, and convergence behavior. As we work on scaling DI-DUCC simulations beyond 18 qubits, we expect that NI-DUCC’s unique advantages will allow it to outperform gradient-based methods like ADAPT-VQE in time cost, facilitating a comprehensive cross-algorithm comparison in future work.

\section{Conclusion}
We have introduced the NI-DUCC-VQE algorithm, designed to construct a disentangled unitary coupled-cluster ansatz, based on a specific set of qubit excitation generators. This method scales linearly with the number of qubits,  and leverages strong symmetry orbitals along with Lie algebraic properties. On a test set of several molecules, we have demonstrated that our method systematically improves accuracy, achieving full convergence, and reaching the Full Configuration Interaction (FCI) reference energy. This is particularly significant at larger internuclear distances, where increased electron correlation leads to multiple electron configurations contributing comparably to the wavefunction. We emphasize that current state-of-the-art fixed ansatz methods in quantum computing for quantum chemistry, such as unitary coupled cluster (UCC) approaches with up to triple (UCCSDT) and quadruple (UsCCSDTQ) fermionic excitations, often fail to capture all correlations, especially as molecular bond lengths move away from equilibrium. In contrast, NI-DUCC not only achieves high accuracy but is also hardware-efficient, making it particularly suitable for NISQ devices. The CNOT count for the NI-DUCC ansatz scales as $\mathcal{O}(knp)$ with its depth $k$, due to the $2n - 2$ generators of MCP that scale linearly, which results in a  relatively shallow circuit depth. Our comparison of CNOT counts between NI-DUCC-VQE and advanced fixed ansatz methods, like COMPASS and COMPACT, reveals that NI-DUCC reduces CNOT usage by more than a factor of 3 for a given accuracy across tested molecules. Remarkably, NI-DUCC surpasses the accuracy of COMPASS and COMPACT with fewer CNOTs, reaching FCI solutions. Comparing NI-DUCC-VQE with ADAPT-VQE, especially QEB-ADAPT-VQE optimized for shallow circuits, we found similar CNOT counts to reach FCI solutions. Unlike ADAPT-VQE, NI-DUCC like COMPACT ansatz, it requires no precircuit measurements for operator gradients. Simulations show that with sufficient layering, NI-DUCC-VQE achieves exponential energy reduction per step, efficiently converging to the FCI solution by avoiding false local minima. This result could suggest NI-DUCC’s potential for addressing barren plateaus in variational algorithms. However, as NI-DUCC layers increase, its parameter count exceeds that of UsCCDTQ, QEB-ADAPT-VQE, COMPACT, and COMPASS, which highights the need for parameter compression strategies like those tools used in COMPASS for instance. This presents a promising direction for further future optimization in NI-DUCC.

Despite these key advantages, we have shown that the NI-DUCC approach comes with a notable cost due to the time complexity involved in the generation of the product group, which is a necessary step to validate the completeness condition of the pool operators. This limitation presently affects large-scale computations beyond 20 qubits. Addressing this issue will require further work. First, we are currently introducing the NI-DUCC approach in the Hyperion-1 GPU-accelerated high performance emulator \cite{adjoua2024} to benefit from further code optimization that will go further than the presently discussed C++ implementation. Second, we are developing a machine learning model to accelerate the verification steps (outlined in Section IV.C (step 4)) by predicting whether a given set of starters meets the inseparability and completeness conditions(explained in details in section 4 of the supplementary Material \cite{thisSuppl}), thereby identifying optimal MCP candidates faster. Nevertheless, despite our current limitations with increasing qubits, NI-DUCC-VQE holds potential for applications on NISQ devices. It can be used for active space selections to target large molecules with frozen core electrons, potentially reaching FCI solutions. Using active space selections within 20 qubits, it can perform accurate calculations with larger chemical basis sets beyond the minimal STO-3G, effectively expanding its applicability to more complex quantum chemistry simulations as qubit resources grow. We aim to tackle the challenges associated with MCP generation as the number of qubits increases, thereby allowing NI-DUCC to advance further with larger basis sets. Additionally, it could be applied to target excited states in both open and closed-shell molecules.

Besides that, our NI-DUCC (Eq. \ref{disentangled1}) is structurally similar to the periodic structure ansatz (see for example Eq. (2) in  \cite{larocca2023theory}), since both rely on layer growth. Periodic structure ansatz have been extensively used to discuss the overparameterization phenomenon in quantum neural networks (QNN) \cite{larocca2023theory}. Overparameterizing a neural network can enhance QNN performance, reduce training and generalization errors \cite{zhang2021understanding,allen2019convergence}, and even result in provable convergence \cite{du2018gradient,brutzkus2017sgd}. This similarity opens the doors to extend NI-DUCC into periodic structure ansatz problems, by integrating them with QNN frameworks in quantum chemistry.
Furthermore, because the NI-DUCC wavefunction relies on Lie algebraic structures, which results in a linear scaling of CNOT gates, it could be beneficial to apply it to other variational quantum algorithms.  The utility of Lie-algebraic structures has been demonstrated in efficient classical simulations across various variational quantum computing algorithms, such as the longitudinal-transverse field Ising model and the quantum approximate optimization algorithm (QAOA) (see  details in Section III of ref. \cite{goh2023lie}). In fact, the dynamic lie (DL) algebraic theory  \cite{larocca2023theory,goh2023lie} has been employed there, to generate sub-algebras of dimension $\mathcal{O}(\text{poly}(n))$, where $n$ denotes the number of qubits \cite{goh2023lie}. It was found that understanding the theoretical connections between the DL sub-algebras and the gradients computations, as demonstrated in \cite{Fontana2024},  has the potential to enhance the trainability of quantum circuits, and exemplify their efficient implementation in numerous paradigmatic tasks in variational quantum computing. In the near future, exploring the integration of the Lie algebraic properties used in our article with dynamic Lie sub-algebra holds promise for advancing the capabilities of Lie algebraic structures in efficiently simulating quantum chemistry problems.

\vspace{1cm}
\noindent\textbf{Acknowledgments}

We thank Huy Binh TRAN for technical support. We thank the HQI initiative for funding S.B postdoctoral position and the PEPR EPIQ - Quantum Software (ANR-22-PETQ-0007, J.-P.P.).  J.-P.P. thanks the European Research Council (ERC) under the European Union’s Horizon 2020 research and innovation program (grant No 810367, project EMC2). Some computations have been performed at IDRIS (Jean Zay) on GENCI Grant no A0150712052 (J.-P.P.).\\

\noindent\textbf{Data and code availability}

Data supporting the findings of this study can be obtained from the corresponding authors upon reasonable request.

\section*{\hspace*{-12.4cm}Additional information}
\label{suppinfo}
\textbf{Supplementary information} The online version contains supplementary material available at \cite{thisSuppl}.

\bibliographystyle{unsrt}
\bibliography{achemso-demo}


\end{document}